# Latency-Sensitive 5G RAN Slicing for Industry 4.0

**Jan García-Morales, (Member, IEEE), M. Carmen Lucas-Estañ, (Member, IEEE), and Javier Gozalvez, (Senior Member, IEEE)**
UWICORE Laboratory, Universidad Miguel Hernández de Elche (UMH), Elche 03202, Spain

Corresponding author: M. Carmen Lucas-Estañ (e-mail: m.lucas@umh.es).

This work has been funded by the European Commission through the FoF-RIA Project AUTOWARE: Wireless Autonomous, Reliable and Resilient Production Operation Architecture for Cognitive Manufacturing (No. 723909), and the Spanish Ministry of Economy, Industry, and Competitiveness, AEI, and FEDER funds (TEC2017-88612-R).

**ABSTRACT** Network slicing is a novel 5G paradigm that exploits the virtualization and softwarization of networks to create different logical network instances over a common network infrastructure. Each instance is tailored for specific Quality of Service (QoS) profiles so that network slicing can simultaneously support several services with diverse requirements. Network slicing can be applied at the Core Network or at the Radio Access Network (RAN). RAN slicing is particularly relevant to support latency-sensitive or time-critical applications since the RAN accounts for a significant part of the end-to-end transmission latency. In this context, this study proposes a novel latency-sensitive 5G RAN slicing solution. The proposal includes schemes to design slices and partition (or allocate) radio resources among slices. These schemes are designed with the objective to satisfy both the rate and latency demands of diverse applications. In particular, this study considers applications with deterministic aperiodic, deterministic periodic and non-deterministic traffic. The latency-sensitive 5G RAN slicing proposal is evaluated in Industry 4.0 scenarios where stringent and/or deterministic latency requirements are common. However, it can be evolved to support other verticals with latency-sensitive or time-critical applications.

## I. INTRODUCTION

5G networks will support the digitalization of key verticals such as manufacturing, automotive, e-health and energy [1]. The digitalization of factories will create smarter and adaptive factories for safer, more energy-efficient and zero-defect production [2]. 5G networks will play a significant role in the development of this Industry 4.0 or Factories of the Future (FoF) vision. The 5G Alliance for Connected Industries and Automation (5G-ACIA) and the 3GPP have already defined Industry 4.0 use cases that can be supported by 5G [3]. This includes use cases related to factory control, monitoring, process automation and maintenance. These use cases include applications with diverse QoS (Quality of Service) requirements in terms of data rate, reliability and latency. These applications can be matched into the 5G service categories: enhanced Mobile Broadband (eMBB), massive Machine Type Communications (mMTC) and ultra-Reliable Low Latency Communications (uRLLC). uRLLC services are of particular relevance to the Industry 4.0 that generally demands low and deterministic latency levels.

5G introduces significant novelties to support the digitalization of verticals, including the Industry 4.0. This includes a New Radio (NR) interface with different numerologies for a flexible use of the radio resources [4]. 5G NR significantly improves the capacity to provide reliable wireless communications with low latency levels. Another important novelty in 5G is the flexibility introduced with the adoption of Software Defined Networking (SDN) and Network Function Virtualization (NFV) technologies. These technologies are fundamental to develop and deploy the concept of Network Slicing (NS) [5]. Network slicing can simultaneously support various services with different QoS requirements over a common physical network infrastructure. To this aim, NS exploits the virtualization and softwarization of networks to create different logical partitions or slices of the common network infrastructure. A slice is formed by a set of network functions, computing, storage, networking and radio resources. Each slice is tailored and configured to support specific applications with distinct QoS requirements. Network Slicing can be applied at the Core Network (CN) or at the Radio Access Network (RAN). To date, most efforts have been devoted to the application of network slicing at the CN (see e.g. [6], [7]). However, it is equally important to address network slicing at the RAN level so that the benefits achieved with network slicing at the CN can positively impact the end-to-end performance. This is particularly critical for latency-sensitive services since the RAN accounts for a relevant part of the end-to-end transmission delay [8]. RAN slicing is in charge of splitting and configuring resources at the RAN level among the slices [9]. This includes defining the slices to adequately serve users (or nodes) with a particular QoS profile, and partitioning (or allocating) the radio resources among the slices [10]. RAN slicing is particularly relevant for latency-sensitive applications since the RAN typical accounts for a large part of the end-to-end service latency [11]. Current RAN slicing solutions are mainly designed with the objective to satisfy the users' bandwidth or rate demands. This approach challenges the capacity to adequately serve latency-sensitive or time-critical applications. These applications are present in many verticals targeted by 5G, including Industry 4.0 where stringent and/or deterministic latency requirements are common. To overcome this limitation, this paper proposes novel RAN slicing schemes for the definition and creation of slices, and the partitioning (or allocation) of radio resources among the slices. The proposals are designed considering both the rate and latency demands of different traffic types. The RAN slicing proposals are evaluated in Industry 4.0 scenarios, and the evaluation demonstrates that the proposals improve the capacity of 5G to satisfy the latency requirements of time-critical Industry 4.0 applications.

The rest of the paper is organized as follows. Section II introduces the concept of RAN slicing in 5G, and Section III reviews related works. Section IV classifies some representative Industry 4.0 use cases and defines their communication requirements. Section V presents our proposal for defining and creating RAN slices. This proposal includes the definition of a novel latency-based slice descriptor that identifies the radio resources necessary to satisfy the latency requirements of different traffic classes. Our proposal is capable to create slices accounting for both rate and latency demands. Section VI presents a novel utility-based partitioning scheme that optimizes the allocation of radio resources to slices based on the requirements of different traffic types and the contributions in Section V. Section VII introduces the reference scheme that is used as a benchmark in this study, and Section VIII describes the evaluation scenario and platform. Section IX presents and analyses the performance achieved with our latency-sensitive 5G RAN slicing solution in Industry 4.0 scenarios. Finally, Section X summarizes the main contributions and conclusions of this study.

## II. RAN SLICING IN 5G

The flexibility that characterizes the 5G New Radio facilitates the deployment of RAN slicing in 5G [12]. 5G NR defines multiple numerologies to support eMBB, uRLLC and mMTC applications with different QoS requirements [4]. Each numerology is characterized by a set of parameters that modify the frame and 5G waveform. Figure 1 compares the 4G and 5G waveforms. 4G (or LTE – Long Term Evolution) defines a fixed slot duration. On the other hand, 5G NR defines different slot durations, and can simultaneously support different numerologies to serve a variety of applications. This flexibility is essential to introduce RAN slicing in 5G.

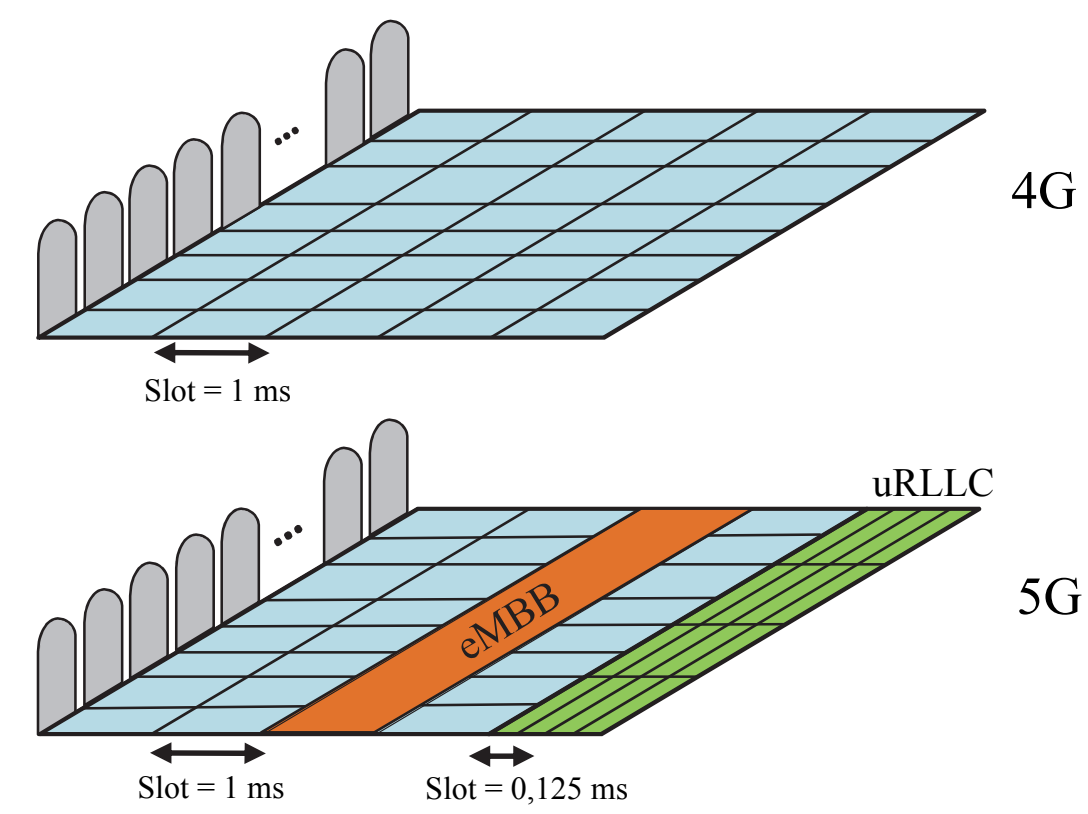

**Figure 1.** Flexible usage of radio resources in 5G NR.

5G NR divides a wideband channel into 10ms frames and 1ms sub-frames. A sub-frame is in turn divided into slots. Slots include 14 consecutive OFDM (Orthogonal Frequency-Division Multiplexing) symbols for a normal cyclic prefix or CP; they include 12 consecutive OFDM symbols for the extended CP. A Resource Block (RB) is the smallest unit of frequency resources that can be allocated to a node. It is defined as 12 consecutive sub-carriers in the frequency domain and one slot in the time domain. Figure 1 illustrates the organization of radio resources into a time/frequency resource grid where the unit is an RB. Each 5G NR numerology $\mu$ modifies the Sub-Carrier Spacing (SCS) $\Delta f$ and the time ($T_{slot}$) duration [13]. Table 1 summarizes some of the main characteristics of the 5G NR numerologies.

**Table 1. 5G NR numerologies [13]**

| Numerology ($\mu$) | $\Delta f$ ($2^{\mu}$·15 [kHz]) | Cyclic prefix | Number of OFDM symbols per slot | $T_{slot}$ [ms] |
|---|---|---|---|---|
| 0 | 15 | Normal | 14 | 1 |
| 1 | 30 | Normal | 14 | 0.5 |
| 2 | 60 | Normal/ Extended | 14/12 | 0.25 |
| 3 | 120 | Normal | 14 | 0.125 |
| 4 | 240 | Normal | 14 | 0.0625 |

RAN slicing can support multiple applications with different QoS requirements thanks to the flexibility introduced in 5G NR and the softwarization and virtualization of the network. This is illustrated in Figure 2 that represents an example where a softwarized and virtualized network can support three RAN slices in a factory environment. The slices share computing, storage and resources at the RAN, but configure differently their radio resources to support eMBB, uRLLC and mMTC applications. For example, slice 1 is configured with shorter time slot durations to support uRLLC applications with low latency requirements. Slice 2 uses a low numerology to support a large number of devices with low bandwidth demands and without strict latency requirements. Slice 3 is configured to support eMBB applications with large bandwidth demands.

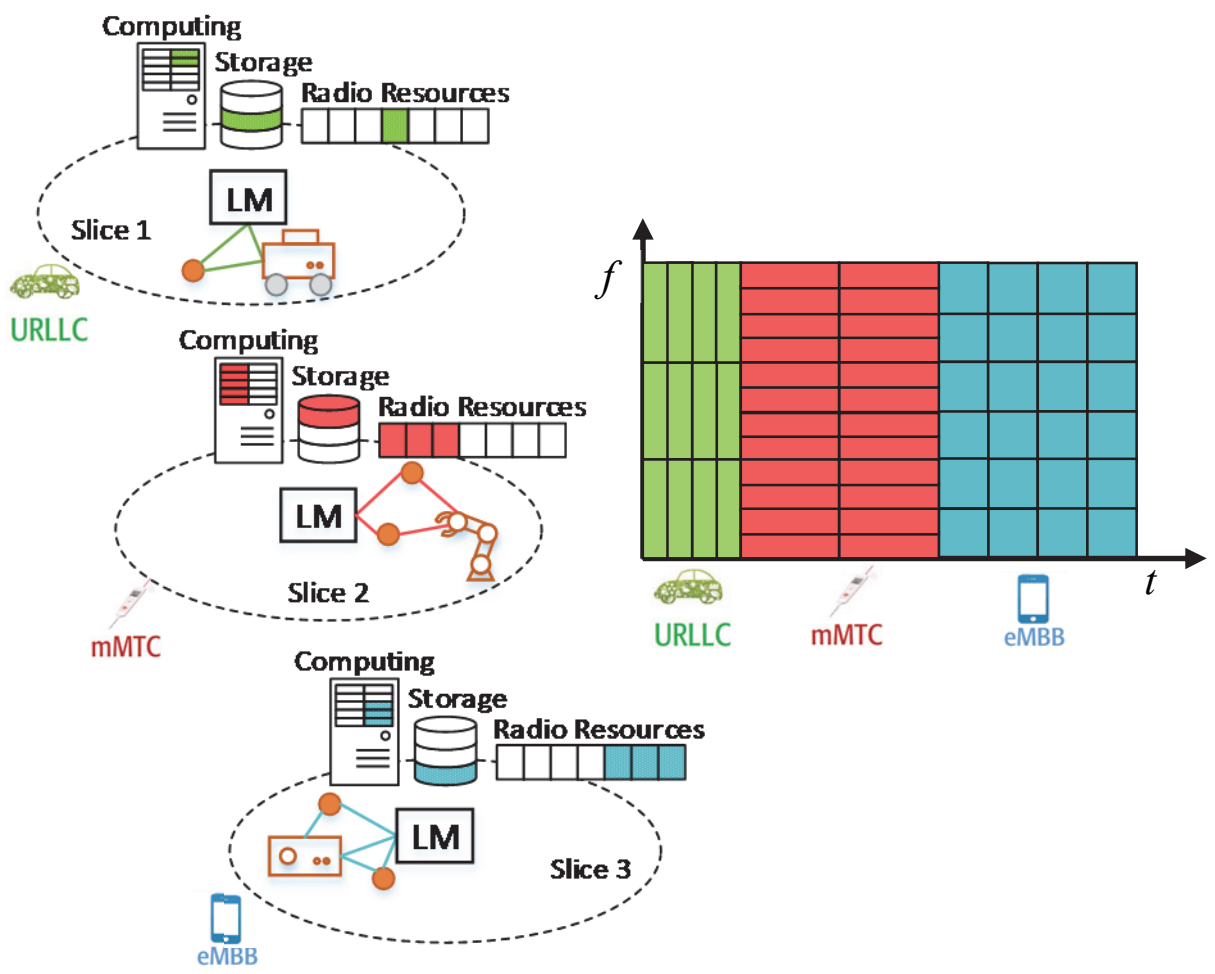

**Figure 2. Illustration of RAN slicing.**

RAN slicing decides how radio resources are configured and allocated to slices in order to support nodes with different QoS requirements [9]. The process to design and create the slices and dynamically allocate the radio resources (or RBs) to the slices is generally referred to as RAN slicing provisioning [14]. The allocation (and configuration of RBs) must be such that the slice can guarantee the QoS requirements of the users it serves. The allocation of RBs to slices is maintained during a time period referred to as allocation window [15]. This period has a duration of $T_w$ slots. The 3GPP defines the lifecycle of slices and the necessary management tasks in [16]. The lifecycle of RAN slices includes the following four main phases that are illustrated in Figure 3:

- Preparation. This phase evaluates the service requirements that will have to be supported by the slices. Based on this analysis, this phase designs the slices. This phase is also in charge of preparing the network environment.
- Commissioning. This phase creates the slices and allocates the RBs among the slices. The 3GPP refers to this process as creation of slices [16] whereas several studies utilize the term partitioning (e.g. [17], [18]). The partitioning scheme is in charge of allocating RBs to slices. The allocation is maintained (at least) for the duration of the allocation window. It can be maintained for longer if conditions do not change.
- Operation. The operation phase includes several management tasks such as supervision and reporting, and resource planning and modification of slices. During this phase, we monitor the performance achieved by the slices and report their main KPIs (Key Performance Indicators). Resource planning computes the usage of the radio resources and requests modifications of the slices if the KPIs are not satisfactory.
- Decommissioning. This phase terminates the slices and releases the RBs. RBs can be allocated to new slices with potentially different configurations.

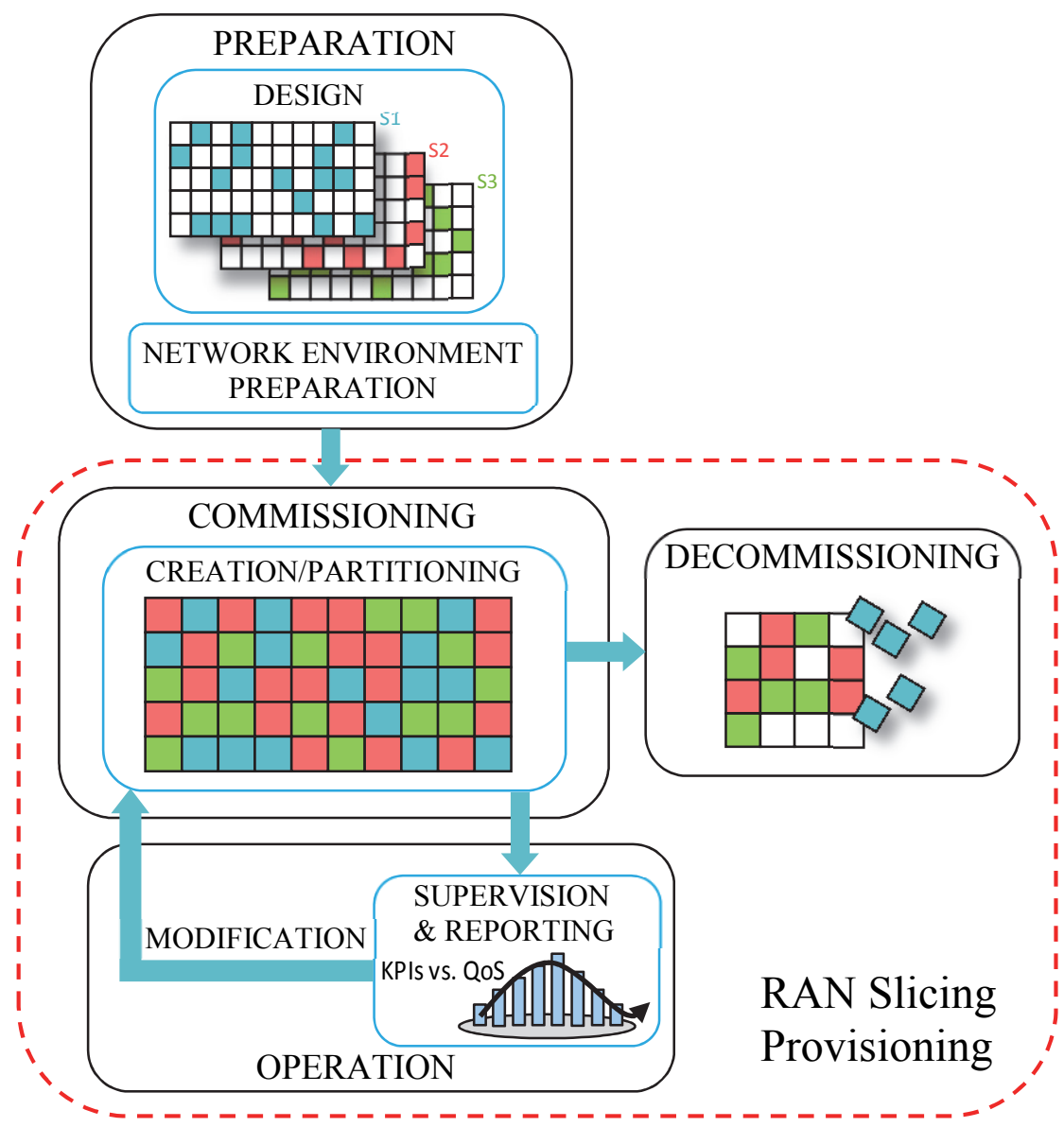

**Figure 3. RAN slicing and lifecycle of slices.**

## III. RELATED WORK

Important efforts have been recently devoted to the development of network slicing in 5G, and in particular of RAN slicing. Authors propose in [19] a general framework for the specification of RAN configuration parameters for the slices. These parameters are referred to as RAN slice descriptors, and are used to characterize the features and resources that define a slice across the radio protocol layers. To date, RAN slices have been generally defined and created considering the number of radio resources necessary to adequately serve users. This is for example the case of the study presented in [20]. The proposal was then extended in [21] to operate using resources from multiple base stations (BSs). These studies take into account the channel quality conditions to decide how many radio resources should be allocated to each slice. This approach is adequate to satisfy bandwidth demands but does not necessarily guarantee any latency requirements. Latency is considered in [22] where authors propose a proactive RAN slicing scheme to support haptic communications. The proposal periodically computes the number of radio resources allocated to each slice. It then uses a dynamic queuing scheme to assign resources to nodes based on their latency requirements. However, these latency requirements are not considered when creating the slices. It is then not possible to guarantee that all nodes will meet their latency requirements with the resources allocated to each slice. Other studies proposed creating slices in mixed traffic scenarios based on bit rate requirements. This is for example the case of [9] that considers a combination of resource-oriented (e.g. occupation of resources) and rate-oriented parameters (e.g. aggregate bit rate) to define and limit the number and characteristics of the resources allocated to each slice. In [23], authors compute the amount of resources necessary per slice based on the aggregate Guaranteed Bit Rate (GBR) requirements of the services. An interesting proposal is presented in [24] to serve elastic and inelastic traffic. Elastic traffic only requires that the average throughput demand is satisfied over a certain time period. On the other hand, inelastic traffic requires that a constant throughput demand is satisfied at all times. The proposal can achieve certain latency levels for inelastic traffic. However, it cannot guarantee any latency requirements since these are not directly embedded in the process to create the slices. To the authors' knowledge, none of the existing studies directly consider latency requirements when creating the slices. This challenges the possibility for RAN slicing to guarantee the stringent latency requirements that characterize latency-sensitive applications such as those found in Industry 4.0.

Once slices are defined, partitioning schemes are used to allocate radio resources to the slices. To this aim, several approaches have been proposed in the literature. The most common one is defining the partitioning process as an optimization problem. For example, [20] and [21] propose a partitioning scheme that is defined as a general integer programing problem. The study in [21] formulates the partitioning process as a Binary Integer Programming (BIP) problem. In [25], authors present a proposal designed to maximize the overall resource utilization (or utility). [22] also proposes a utility-based partitioning strategy based on a reinforcement learning. A dynamic partitioning process is defined in [26] where authors introduce the concept of a Slice Broker. The broker initially reserves an amount of resources per slice, and monitors the traffic per slice. It increases the allocation of resources per slice if necessary. The challenge with this approach is that it can incur in some delay until the broker allocates the adequate number of resources to each slice. A Markovian approach with slice-aware admission control is proposed in [23] for sharing resources in multi-tenant scenarios with diverse guaranteed bit rate services. The proposal in [27] focuses on reliability, and formulates a risk-sensitive partitioning optimization problem to satisfy the reliability requirements of eMBB and uRLLC services. An alternative to optimization problems is the design of partitioning schemes using game theory. This is for example the case of the study in [28] that uses bankruptcy theory for the allocation of resources to slices. The resource utilization is improved using cooperative sharing. [29] also proposes a RAN slicing game, and shows it is possible to reach a Nash equilibrium under certain conditions. The use of game theory is interesting but challenging when considering latency-sensitive use cases such as those found in Industry 4.0.

The review of the state of the art has shown that current solutions for the creation of slices and the partitioning or allocation of resources to slices do not directly consider latency in their design. This limits the possibility for RAN slicing to adequately support latency-sensitive or time-critical applications. These applications are particularly relevant in Industry 4.0 scenarios where stringent and/or deterministic latency requirements are common. To overcome this limitation, this paper proposes novel schemes for the creation of slices and the partitioning (or allocation) of the radio resources to slices. These schemes differentiate between traffic types, and directly embed in their design the rate and latency requirements of each traffic class. This study is conducted in the framework of the European H2020 AUTOWARE project. The project focuses on the design of wireless solutions for Industry 4.0. We then present relevant Industry 4.0 use cases and communication requirements before describing our latency-sensitive RAN slicing proposals.

## IV. INDUSTRY 4.0 USE CASES AND COMMUNICATION REQUIREMENTS

The Industry 4.0 (or Factories of the Future) paradigm envisions a series of changes to transform the current relatively static and long-lasting production facilities in highly flexible connected and digitalized factories. Future smart factories will need to be more flexible and integrate more efficiently mobile robots, reconfigurable machinery and mobile industrial applications [3]. This requires a higher integration of wireless communication in factories, and 5G is certainly an important enabler for the Industry 4.0 [30]. The 5G Alliance for Connected Industries and Automation (5G-ACIA) and the 3GPP have identified in [3] and [30] the Industry 4.0 use cases and applications. The use cases are related to different application areas, such as process and factory automation, and logistics warehousing, monitoring and maintenance, among others. The use cases have different communication requirements defined in terms of data rates, latency, reliability or availability among others. The use cases are classified in [3] into three different traffic classes: deterministic periodic, deterministic aperiodic and non-deterministic (periodic or aperiodic). Deterministic periodic traffic is generated periodically and must be received within a given time deadline. Deterministic traffic is characterized by a maximum latency that depends on the supported use case. Deterministic periodic traffic is the most common industry traffic class [3]. For example, it relates to use cases such as motion control, control to control communication, mobile robot communication, and process automation among others. Deterministic aperiodic traffic stands for traffic that is not generated periodically, but when packets are generated they must also be received with a given deadline. Deterministic aperiodic traffic is characteristic of event-driven use cases where a transmission is triggered when specific events occur. These events can be activated, for example, when: 1) a temperature, pressure or level exceeds or falls below predefined thresholds (process events), 2) sensors detect malfunctions or errors of devices or modules, 3) or based on information that indicates necessary maintenance work to prevent failures (maintenance events). Deterministic aperiodic traffic is, for example, characteristic of uses cases related to control panels with safety functions and process automation. Finally, non-deterministic traffic is traffic (periodic or aperiodic) that does not have a time deadline by which it must be received. Non-deterministic traffic is characteristic of applications that for example require software updates or file downloads among others. These applications can be found in use cases such as motion control, safety panels or process automation among others. The main characteristics and communication requirements of some selected representative Industry 4.0 use cases are described below. We have selected use cases and applications for each traffic class. Their requirements are summarized in Table 2 where use cases and applications are grouped based on their traffic class. A detailed analysis of all use cases and their requirements can be found in [3].

**Table 2. Industry 4.0 use cases and applications [3]**

| Traffic class | Use Case | Application | Latency | Payload | Data Rate | # Nodes |
|---|---|---|---|---|---|---|
| Deterministic periodic traffic | Motion control | Printing machine | 2 ms | 20 bytes | - | 20-100 |
| | Control to Control | Machines coordination | 4 - 10 ms | 1 Kbytes | - | 5 - 10 |
| | Mobile robots | Cooperative control | 1 ms | 40 - 250 bytes | - | <100 |
| Deterministic aperiodic traffic | Safety panels | Emergency stops | <4 ms | 40 - 250 bytes | - | <100 |
| | P.A. – closed-loop control | Closed-loop control events | <10 ms | 10 bytes | - | 10 - 1000 |
| | P.A. – plan asset management | Failure alarms | <50 ms | 10 - 100 bytes | - | <10.000/Km$^2$ |
| Non-deterministic traffic | Motion control | Software/firmware updates | - | - | > 1 Mbps | <100 |
| | Safety panels | User interaction | - | - | > 5 Mbps | <4 |
| | P.A. – plan asset management | Assets software updates | - | - | > 1 Mbps | <10.000/Km$^2$ |

- *Motion control*: A motion control system is responsible for controlling moving and/or rotating parts of machines (e.g. printing machines, machine tools or packaging machines). Motion control generates periodic traffic with deterministic and stringent latency requirements. This use case can also require non-real-time data related for example to software/firmware updates or maintenance information. This use case is hence included in two traffic classes in Table 2.
- *Control-to-control communication*: This use case relates to the communication between different industrial controllers. Such communication can be necessary to connect, for example, individual machines that are used in an assembly line for fulfilling a common task. It can also be required to synchronize and exchange real-time data between different controllers in large machines (e.g. newspaper printing machines). Control-to-control communication typically generates periodic traffic with deterministic and near real-time latency requirements.
- *Mobile robots*: A mobile robot is a programmable machine able to fulfil a large variety of tasks usually following programmed paths. Mobile robots are normally controlled or monitored from a guidance control system. A deterministic and periodic communication between the robot and the control system is usually required. Other types of traffic might also be demanded depending on the specific application supported by the mobile robot.
- *Mobile control panels with safety functions (safety panels)*: Control panels are mainly used for configuring, monitoring, and controlling machines, robots, or production lines. Safety control panels are also typically equipped with an emergency stop button. This use case requires the transmission of non-critical data (non-deterministic traffic) for the configuration, monitoring, and maintenance of the machines. It also requires the transmission of highly-critical and unpredictable safety data with stringent latency requirements (deterministic aperiodic traffic) when pressing the emergency stop button.
- *Process automation (P.A.) – closed-loop control*: In this use case, several sensors are installed in a plant and each sensor makes continuous measurements. The sensed data is transmitted to a controller that acts on certain actuators. The latency and determinism in this use case are crucial. Closed-loop control produces periodic and aperiodic traffic with strict latency requirements (i.e. deterministic traffic). The traffic is aperiodic if for example the sensor only transmits data when a certain threshold is exceeded. It will be periodic if the sensed data must be periodically transmitted to maintain the industrial process active.

- *Process automation (P.A.) – plan asset management*: In this use case, sensors collect data about assets. This data must be transmitted for storage and processed within a defined time interval (deterministic aperiodic traffic). This data is used to continuously diagnose assets and components, and be able to detect (and even predict) any possible degradation. If a failure or degradation is detected, an event is transmitted immediately. This use case can also include remote software updates when, for example, it is necessary to adapt components to changing conditions.

## V. DESIGN OF RAN SLICES

A critical step in RAN slicing is the design of the slices. This is done before the creation of a slice in the preparation phase as defined by the 3GPP in [16]. The slices must be designed to satisfy the communication requirements of the services to be supported by the slices. To date, most proposals define slices in terms of the number of radio resources required to satisfy a bandwidth or rate demand. However, this descriptor does not account for latency requirements that are fundamental in certain 5G-enabled verticals such as the Industry 4.0. This study addresses this limitation, and proposes to utilize two descriptors to define the RAN slices. The first one is the number of radio resources (or RBs) needed to satisfy the services' bandwidth or rate requirements. This descriptor is the most commonly used to date, and is referred to in this paper as the size of the slice. The second descriptor is a novel latency-based descriptor proposed by the authors. It accounts for the latency requirements of the supported services, and is referred to as the shape of the slice. The shape of a slice is defined by the slots over which the number of RBs that define the size of the slice must be reserved. As a result, the shape of a slice indicates the relative position of the RBs that must be allocated to a slice in order to satisfy the latency requirements of the traffic supported by the slice. This section analytically estimates the size and shape of slices for the three traffic classes that characterize Industry 4.0 use cases and applications. Examples of size and shape are also provided for each traffic class.

### A. NON-DETERMINISTIC TRAFFIC

A slice $s_i$ is created to support a group $G_i$ of nodes with similar QoS requirements. In the case of applications with non-deterministic traffic, nodes in $G_i$ demand a minimum data rate $R_i$ (see examples in Table 2). The size of a slice is defined as the number of RBs that must be reserved for a slice $s_i$ (within the allocation window) to satisfy the data rate $R_i$ demanded by each node. Following [20], we define $R_u^{\text{eff}}$ as the effective transmission rate or throughput that node $u$ will experience per assigned RB. This throughput is a function of the experienced Signal-to-Interference-plus-Noise Ratio (SINR) and the reliability required by the application. $R_u^{\text{eff}}$ is defined as:

$$R_u^{\text{eff}}(SINR_u) = \frac{TBS(SINR_u)}{T_w}(1 - BLER) \quad (1)$$

where $SINR_u$ is the SINR experienced by node $u$ on a RB. $TBS(SINR_u)$ represents the Transport Block Size (TBS in bits) that can be transmitted over a RB. The TBS is a function of the SINR since the SINR establishes the Modulation and Coding Scheme (MCS) that can be used for a transmission. MCSs with higher error correction capabilities can operate with lower SINR levels but transmit fewer bits per RB. The MCS is selected based on the experienced SINR and the BLER necessary to deliver the data. We select the MCS with the larger TBS that guarantees the target BLER for the experienced SINR. The MCS is selected using the lookup table specified in [31]. This lookup table maps the SINR to the MCS necessary to guarantee a target BLER. Using this lookup table, we obtain the value of $TBS(SINR_u)$ for the $SINR_u$ experienced by node $u$. Table 3 shows the MCSs and $TBS(SINR_u)$ for different values of the SINR based on the lookup table in [31]. We consider a target BLER equal to $10^{-5}$ following [32].

**Table 3. Lookup table [31]**

| $SINR_u$ [dB] | MCS Index | $TBS(SINR_u)$ [bits] |
|---|---|---|
| ≥-0.4167 | 0 | 16 |
| ≥ 1.0417 | 1 | 24 |
| ≥ 1.6667 | 2 | 32 |
| ≥ 2.9167 | 3 | 40 |
| ≥ 3.5417 | 4 | 56 |
| ≥ 5.0000 | 5 | 72 |
| ≥ 5.6250 | 6 | 88 |
| ≥ 7.0833 | 7 | 104 |
| ≥ 7.9167 | 8 | 120 |
| ≥ 8.7500 | 9 | 136 |
| ≥10.8333 | 10 | 144 |
| ≥11.6667 | 11 | 144 |
| ≥12.9167 | 12 | 176 |
| ≥13.3333 | 13 | 208 |
| ≥14.5000 | 14 | 224 |
| ≥15.0000 | 15 | 256 |
| ≥15.8333 | 16 | 280 |
| ≥15.9167 | 17 | 328 |
| ≥16.0000 | 18 | 336 |
| ≥16.2917 | 19 | 376 |
| ≥16.8750 | 20 | 408 |
| ≥18.6667 | 21 | 408 |
| ≥19.7917 | 22 | 440 |
| ≥20.4167 | 23 | 488 |
| ≥21.0417 | 24 | 520 |
| ≥21.4167 | 25 | 552 |
| ≥22.7083 | 26 | 584 |
| ≥24.5000 | 27 | 616 |
| ≥25.8333 | 28 | 712 |

The number of RBs required by node $u$ to achieve $R_i$ can be expressed as follows:

$$J_u(SINR_u) = \left\lceil \frac{R_i}{R_u^{\text{eff}}(SINR_u)} \right\rceil \quad (2)$$

where $\lceil x \rceil$ denotes the ceil operator.

A slice should serve a group of nodes with similar QoS requirements. The size of a slice $s_i$ ($K_i^{\text{size}}$) created to serve $M$ nodes is then:

$$K_i^{\text{size}} = \sum_{u=1}^{M} J_u(SINR_u), \quad \forall s_i \in S_n \tag{3}$$

where $S_n$ represents the set of slices that support non-deterministic traffic applications. $SINR_u$ in (3) is not an instantaneous SINR level but an average one. This is the case because an instantaneous value does not adequately reflect the SINR that nodes can experience during the complete allocation window. To compute the average SINR, nodes measure the experienced SINR every 1ms, and store the measurements of the last second. $SINR_u$ is the average SINR value experienced by node $u$ during the last second.

Non-deterministic traffic does not define a latency deadline by which data must be received. The average requested data rate must then only be satisfied within the allocation window. In this context, any RB within the allocation window can be selected as one of the $K_i^{\text{size}}$ RBs that form the slice. The shape of the slice includes then all the slots in the allocation window and the following condition must be satisfied:

$$\sum_{t=1}^{T_w} L_{i,t} = K_i^{\text{size}} \tag{4}$$

where $L_{i,t}$ is the amount of RBs allocated to slice $s_i$ in slot $t$.

Figure 4 illustrates an example of the size and shape of a slice for non-deterministic traffic. The grid represents the RBs in the time and frequency domains. The example represents the case where an application requires a slice size of four RBs. The four selected RBs can be part of any of the slots within the allocation window for this traffic class.

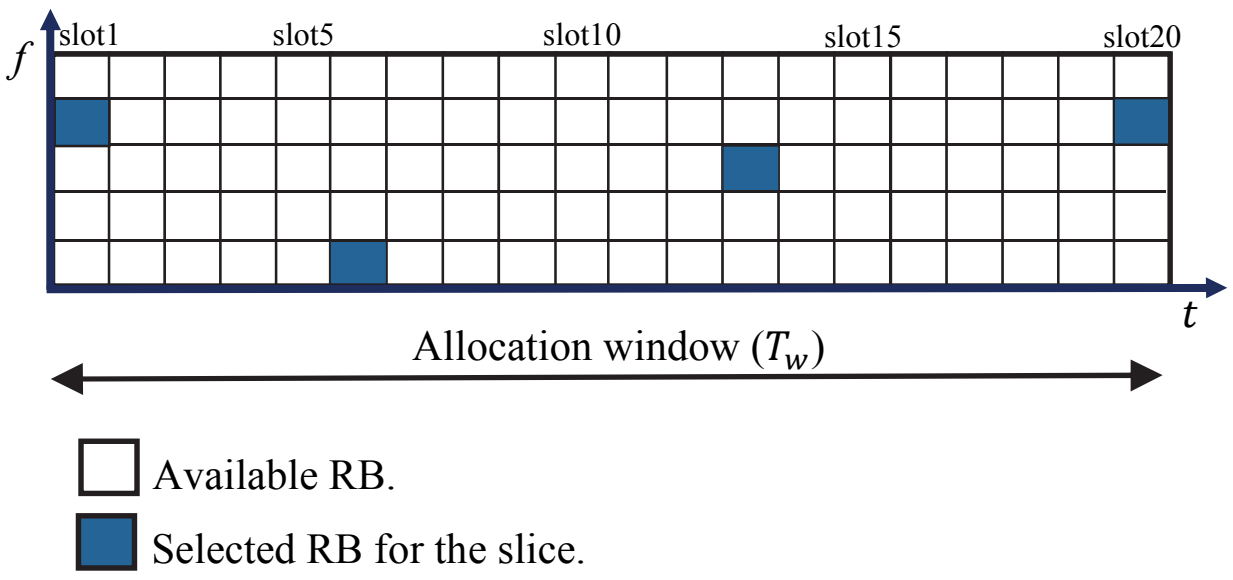

**Figure 4. Size and shape of a slice for non-deterministic traffic. Example with an allocation window of 20 slots and $K_i^{\text{size}}$ = 4 RBs.**

### B. DETERMINISTIC PERIODIC TRAFFIC

Applications with deterministic periodic traffic generate packets periodically, and packets must be received before a maximum latency deadline. We consider that a slice $s_i$ is created to support a group $G_i$ of nodes that generate deterministic periodic traffic with similar QoS requirements. In this case, $G_i$ is characterized by a transmission period $T_p^i$, a payload of $b_i$ bits, and a deadline $D_i$. The size of a slice is then the number of RBs within the transmission period $T_p^i$ that must be reserved for a slice $s_i$ in order to satisfy the rate required by the nodes. The data rate $R_i$ (in bps) required by a node included in $G_i$ to transmit a payload of $b_i$ bits before $D_i$ is:

$$R_i = \frac{b_i}{D_i} \tag{5}$$

The effective transmission rate or throughput $R_u^{\text{eff}}$ that a node $u$ will experience per assigned RB can be expressed as:

$$R_u^{\text{eff}}(SINR_u) = \frac{TBS(SINR_u)}{D_i}(1 - BLER) \tag{6}$$

Eq. (6) is similar to (1) for non-deterministic traffic except that $T_w$ is replaced by $D_i$ in (6). We compute then the number of RBs required by node $u$ to transmit $b_i$ bits before $D_i$ using (2), (5) and (6). The size of a slice $s_i$ ($K_i^{\text{size}}$) created to serve $M$ nodes during a transmission period $T_p^i$ can then be expressed as:

$$K_i^{size} = \sum_{u=1}^{M} J_u(SINR_u), \quad \forall s_i \in S_p \tag{7}$$

where $S_p$ represents the set of slices that support deterministic periodic applications. The SINR level $SINR_u$ is also an average value, and is computed like in the case of non-deterministic traffic.

The shape of the slice identifies the slots within the transmission period $T_p^i$ that must contain the $K_i^{\text{size}}$ RBs that have to be reserved to guarantee the latency requirements demanded by nodes in $G_i$. We must guarantee that all $K_i^{\text{size}}$ RBs are available between the time a new packet is generated and the latency deadline $D_i$. It is possible to estimate the time at which packets are generated in the case of deterministic periodic traffic. We define $L_{i,t}$ as the number of RBs allocated to slice $s_i$ in slot $t$. To meet the latency deadline $D_i$, the slice must be created so that:

$$\sum_{t=t_z}^{t_z+D_i-1} L_{i,t} = K_i^{\text{size}}, \qquad \forall t_z \in T_0 \tag{8}$$

where $D_i$ is expressed as an integer number of slots, and $T_0 = \{t_z \mid t_z = t_0 + zT_p^i, \forall z \in \{0,1,\dots,\lfloor T_w/T_p^i \rfloor - 1\}\}$, $t_0$ is the time slot at which the first transmission starts, and $t_z$ is the time at which packet $z$+1 is generated.

Figure 5 illustrates an example of the size and shape of a slice for deterministic periodic traffic. For illustration purposes, we consider that the slice only supports a node. The example represents the case where the node requires a slice with a size of four RBs in each transmission period. The $K_i^{\text{size}}$ RBs (or size of the slice) must be contained within a time window of length $D_i$ from the start of every transmission at $t_z$. Figure 5 represents different examples of shapes with the same value of $K_i^{\text{size}}$. They all guarantee the availability of $K_i^{\text{size}}$ RBs before $D_i$ from the start of each transmission. These examples illustrate the relevance of the latency descriptor proposed by the authors.

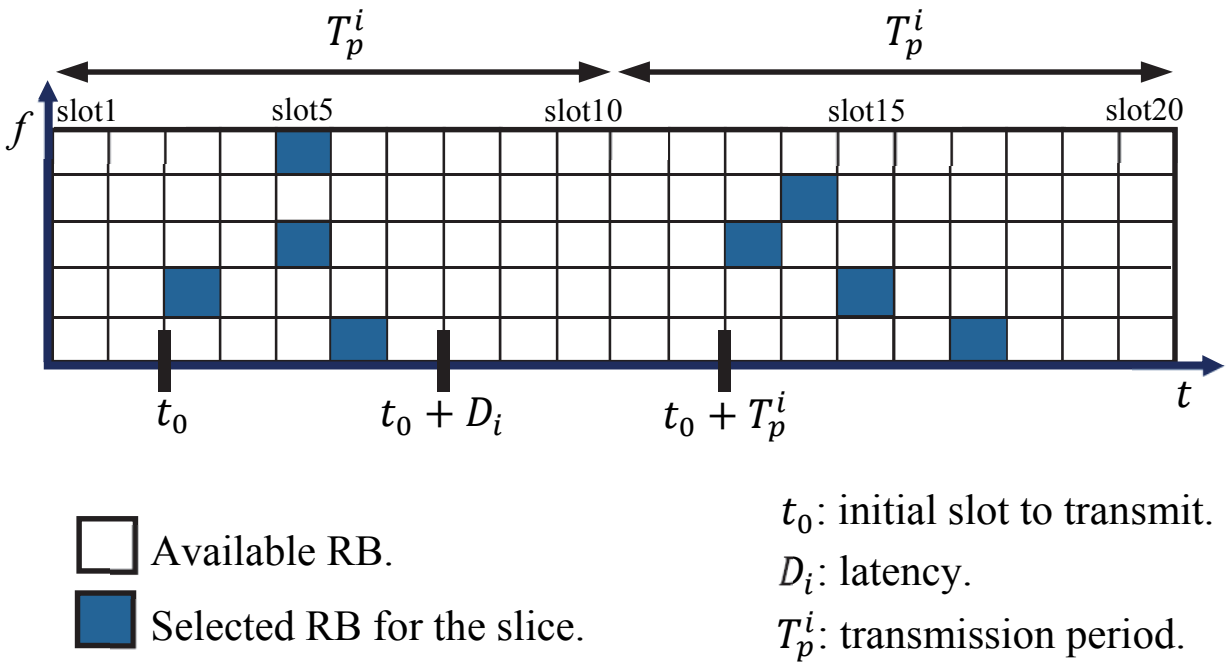

**Figure 5. Size and shape of a slice for deterministic periodic traffic. Example with an allocation window of 20 slots and $K_i^{\text{size}}$ = 4 RBs.**

### C. DETERMINISTIC APERIODIC TRAFFIC

Applications with deterministic aperiodic (or sporadic) traffic can generate packets with a given payload at any point in time. The packet generation rate is not periodic, and it is not possible to predict when packets will have to be transmitted. However, once a packet is generated it must be delivered before a latency deadline $D_i$ with reliability $P_{rel}$. Following [12], reliability is defined as the percentage of packets that are successfully delivered before the latency deadline. Similarly to previous traffic classes, we also define $s_i$ as a slice created to support a group $G_i$ of nodes with deterministic aperiodic traffic and similar QoS requirements. Nodes in $G_i$ are here characterized by a payload of $b_i$ bits, and a deadline $D_i$. The size of a slice is then the number of RBs that must be reserved from the time a packet is generated until the transmission deadline. The number of RBs must satisfy the rate required by the deterministic aperiodic application and guarantee the correct reception of packets with probability $P_{rel}$.

One approach to compute the size of a slice for deterministic aperiodic traffic would be to reserve for each node $u$ served by the slice $J_u(SINR_u)$ RBs within any time period equal to the latency deadline $D_i$. This would ensure that all nodes have the necessary resources to satisfy their rate demand $R_i$ and latency deadline $D_i$ independently of when packets are generated. However, it would imply a very inefficient use of resources since slices would be over-dimensioned. The 5G NR standard introduces the possibility that nodes share resources [33]. This is an interesting option to support deterministic aperiodic traffic and utilize efficiently the radio resources. In this case, nodes have to contend for the use of RBs anytime they have a packet to transmit. This can result in packet collisions. These collisions can be reduced if nodes randomly select their RBs among the available ones [34]. We adopt this proposal to define and create the slices that serve applications with deterministic aperiodic traffic.

Let's consider that a slice $s_i$ should serve $M$ nodes that generate deterministic aperiodic traffic. The nodes share the RBs assigned to the slice. Following [34], nodes randomly select their RBs among the $k$ available ones when they have a packet to transmit. Without loss of generality, we assume that each node generates packets following a Poisson distribution with exponential inter-arrival time [34]. The average packet inter-arrival time is equal to $1/\lambda$. $\lambda$ is the average number of packets generated per second. The probability $P_p$ that a node generates one or more packets in a time interval equal to $T_{slot}$ is:

$$P_p = 1 - \exp(-T_{slot}\,\lambda) \tag{9}$$

The probability $P_c$ that a packet collides is computed in [34] and is expressed as follows:

$$P_c = 1 - \left(\frac{k - \bar{J}P_p}{k}\right)^{M-1} \tag{10}$$

where $k$ is the number of available RBs, and $\bar{J}$ represents the average number of RBs required per node. $\bar{J}$ can be computed as:

$$\bar{J} = \frac{1}{M}\sum_{u=1}^{M} J_u(SINR_u) \tag{11}$$

where $J_u(SINR_u)$ is obtained using (2). $SINR_u$ is again the average SINR measured by node $u$ over the last second.

We need to guarantee a reliability $P_{rel} \geq 1\text{-}P_c$. We can then estimate the minimum number $k$ of RBs necessary to satisfy $P_{rel}$ when $M$ nodes share the RBs as:

$$k = \frac{\bar{J}\,P_p}{1 - P_{rel}^{\frac{1}{M-1}}} \tag{12}$$

Eq. (12) identifies the number of RBs necessary to satisfy the requirements of $M$ nodes that generate deterministic aperiodic traffic and share the RBs of a slice. The number of dedicated RBs necessary to serve these $M$ nodes is equal to $\sum_{u=1}^{M} J_u(SINR_u)$. The size of the slice within a time window equal to $D_i$ is then:

$$K_i^{\text{size}} = \min\left(k\,, \sum_{u=1}^{M} J_u(SINR_u)\right), \quad \forall s_i \in S_a \tag{13}$$

where $S_a$ represents the set of slices that support deterministic aperiodic applications.

The shape of the slice must be so that any node served by the slice can access at any point in time $l$ the necessary RBs to satisfy the $R_i$ demand and transmit its packets before $D_i$ with reliability $P_{\text{rel}}$. The shape of the slice identifies the slots within the allocation window over which the $K_i^{\text{size}}$ RBs must be reserved. The following condition must be satisfied for a slice to support deterministic aperiodic traffic:

$$\sum_{t=l}^{l+D_i-1} L_{i,t} = K_i^{\text{size}}, \qquad \forall l \in [1, T_w] \tag{14}$$

$L_{i,t}$ is the number of RBs allocated to slice $s_i$ in slot $t$. The expression in (14) guarantees the availability of $K_i^{\text{size}}$ RBs within a time window $D_i$ from any time $l$ at which a packet is generated.

Figure 6 illustrates an example of the size and shape of a slice for deterministic aperiodic traffic. This example represents the case where applications require slices with a size of 4 RBs and a latency requirement $D_i$ of 5 slots. The 4 RBs must be available within 5 slots from any time $l$ at which a packet is generated. This is actually the case for the slice illustrated in Figure 6. It is possible to visually verify that for any possible $l$ there are always 4 RBs in the slice within the 5 slots from $l$.

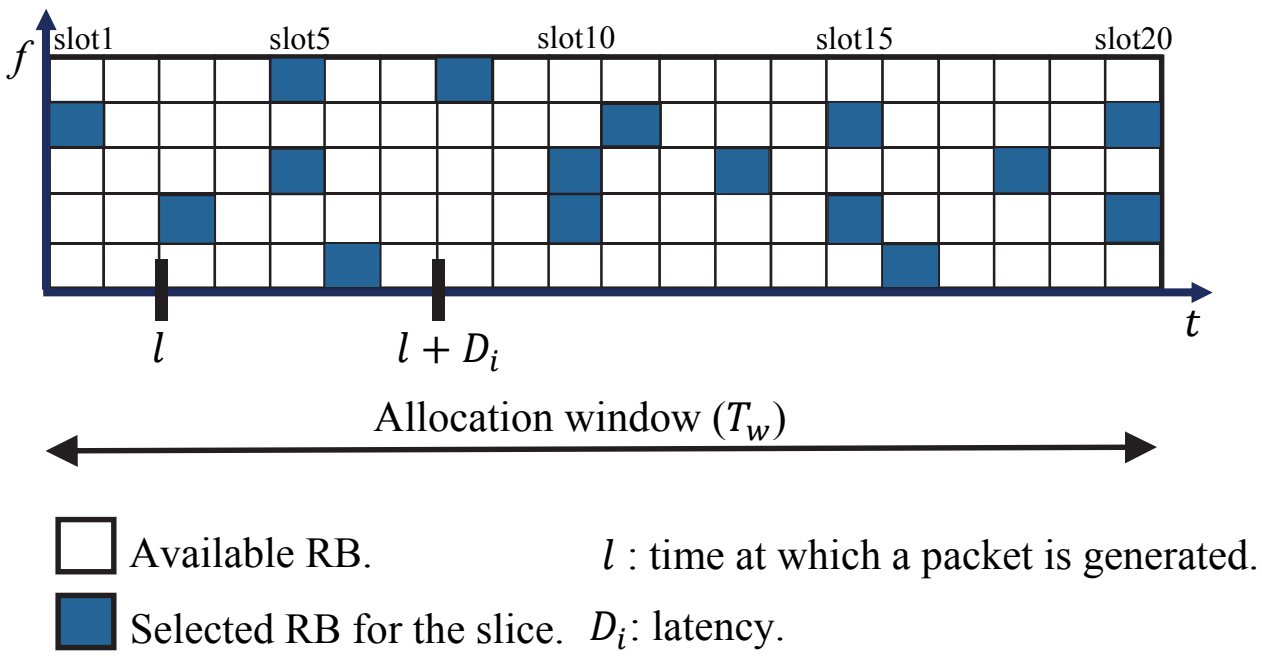

**Figure 6. Size and shape of a slice for deterministic aperiodic traffic. Example with an allocation window of 20 slots and $K_i^{\text{size}}$ = 4.**

## VI. PARTITIONING OF RESOURCES

This section presents a novel partitioning scheme that allocates RBs to slices based on their traffic class, size and shape. The partition (or allocation) of RBs among the slices is executed during the creation of the slice in the commissioning phase as defined by the 3GPP in [16]. We consider that a 5G NR network deployed in a factory needs to create different RAN slices to support a variety of applications. Each slice serves a group of nodes with similar QoS requirements. Each slice is then characterized by a specific size and shape following Section V. The objective of the partitioning proposal is to maximize the number of satisfied slices that receive the RBs necessary to match their size and shape. When all slices are satisfied, the partitioning scheme distributes any available RBs among the slices to improve the QoS.

The partitioning scheme distributes the RBs for the duration of the allocation window. We consider that this duration is equal to $T_w$ slots and that there are $N_{RB}$ RBs per slot. Let's define $S$ as the set of slices to be created. $S_p$, $S_a$, and $S_n$ are the sets of slices supporting deterministic periodic, deterministic aperiodic and non-deterministic traffic respectively. The following relation is then valid: $S$=$S_p$∪$S_a$∪$S_n$.

Non-deterministic traffic has no latency requirements. In this case, the RBs assigned to slices $s_i \in S_n$ are reserved for the complete duration of the allocation window. A slice $s_i \in S_n$ is satisfied if the partitioning scheme assigns a number $K_i$ of RBs to $s_i$ within the allocation window higher than $K_i^{\text{size}}$ following (3). We define $H_i(K_i)$ for slice $s_i$ as:

$$H_i(K_i) = \begin{cases} 0, & K_i < K_i^{\text{size}} \\ 1, & K_i \geq K_i^{\text{size}} \end{cases} \tag{15}$$

$H_i(K_i)$ is then equal to 1 if $s_i$ receives sufficient RBs to satisfy its size requirement, and equal to 0 otherwise. The size $K_i^{\text{size}}$ of a slice $s_i \in S_n$ is defined in (3). We can then maximize the number of satisfied RAN slices that support non-deterministic traffic by solving the following optimization problem:

$$\text{o.f.: max} \sum_{\forall s_i \in S_n} H_i(K_i) \tag{16}$$

$$\text{s.t.:} \quad \sum_{t=1}^{T_w} L_{i,t} = K_i, \qquad \forall s_i \in S_n \tag{17}$$

where $L_{i,t}$ is the amount of RBs allocated to slice $s_i$ in slot $t$, and the constraint in (17) defines the shape requirement. Non-deterministic traffic does not have latency requirements. Consequently, the shape requirement only establishes that the demanded $K_i^{\text{size}}$ RBs must be assigned within the allocation window.

Deterministic periodic traffic generates packets periodically. The period is defined as the transmission period and is denoted as $T_p^i$ for slice $s_i \in S_p$. The value of $T_p^i$ can vary for each $s_i \in S_p$. The size $K_i^{\text{size}}$ of a slice $s_i \in S_p$ is the number of RBs that the slice needs within each transmission period $T_p^i$. This size is defined in (7). We can maximize the number of satisfied RAN slices for deterministic periodic traffic by solving the following optimization problem:

$$\text{o.f.: max} \sum_{\forall s_i \in S_p} \left[ \frac{1}{\lceil T_w / T_p^i \rceil} \sum_{\forall t_z \in T_0} H_i(K_i) \right] \tag{18}$$

$$\text{s.t.:} \quad \sum_{t=t_z}^{t_z+D_i-1} L_{i,t} = K_i, \ \forall t_z \in T_0 \,\&\, t_z \leq T_w\text{-}(D_i\text{-}1), \forall s_i \in S_p \tag{19}$$

$$\sum_{t=t_z}^{T_w} L_{i,t} + \sum_{t=1}^{\text{mod}\left(\frac{t_z+D_i-1}{T_w}\right)} L_{i,t} = K_i, \tag{20}$$
$$\forall t_z \in T_0 \,\&\, t_z > T_w\text{-}(D_i\text{-}1), \forall s_i \in S_p$$

where the constraints in (19) and (20) are the shape requirements for slices supporting deterministic periodic traffic. Constraint in (19) relates to transmissions that start and end in the same allocation window considering the maximum latency requirements. Constraint in (20) relates to transmissions that start in an allocation window but may end in the following allocation window considering their latency requirements. These constraints specify that the $K_i^{\text{size}}$ RBs must be available within $D_i$ from the time $t_z$ the packet is generated to satisfy the latency demand of deterministic traffic. Eq. (20) then guarantees that the latency requirements are guaranteed beyond the boundary of the allocation windows. Eq. (18)-(20) specify that the demand characterizing the size and shape of each slice must be satisfied for all the transmission periods that are included within an allocation window. The factor $1/(\lfloor T_w/T_p^i \rfloor)$ is introduced in (18) so that all slices $s_i \in S_p$ have the same weight in the resolution of the optimization problem.

Deterministic aperiodic traffic can generate packets at any point in time within the allocation window. Once a packet is generated, it must be delivered before a given latency deadline. The size $K_i^{\text{size}}$ of a slice is then the number of RBs that must be reserved in a time window that can start at any time instant (the traffic is aperiodic) and has a duration equal to the latency deadline. We must then ensure that there are $K_i^{\text{size}}$ RBs reserved for each slice $s_i \in S_a$ within each time period of duration $D_i$ in the allocation window. The size $K_i^{\text{size}}$ for slices serving applications with deterministic aperiodic traffic is defined in (13). We can maximize the number of satisfied RAN slices for deterministic aperiodic traffic by solving the following optimization problem:

$$\text{o.f.:} \max \sum_{\forall s_i \in S_a} \left[ \frac{1}{T_w} \sum_{l=1}^{T_w} H_i(K_i) \right] \tag{21}$$

$$\text{s.t.:} \sum_{t=l}^{l+D_i-1} L_{i,t} = K_i, \;\; \forall l \in [1, T_w\text{-}(D_i\text{-}1)], \forall s_i \in S_a \tag{22}$$

$$\sum_{t=l}^{T_w} L_{i,t} + \sum_{t=1}^{\text{mod}\left(\frac{l+D_i-1}{T_w}\right)} L_{i,t} = K_i, \quad \forall l \in [T_w\text{-}(D_i\text{-}1)+1, T_w], \forall s_i \in S_a \tag{23}$$

where the constraints in (22) and (23) are the shape requirements for slices supporting deterministic aperiodic traffic. A sporadic transmission can be generated at any time $l \in [1, T_w]$. Eq. (22) and (23) account then for all transmission possibilities within the allocation window for each slice $s_i \in S_a$. Eq. (22) accounts for transmissions that start and end in the same allocation window considering the maximum latency requirements. Eq. (23) accounts for transmissions that start in an allocation window and that may end in the following allocation window considering the maximum latency requirements. Eq. (23) guarantees that the latency requirements are satisfied beyond the boundary of the allocation window. The factor $1/T_w$ in (21) is introduced so that all slices $s_i \in S_a$ have the same weight in the resolution of the optimization problem.

The objective of the proposed partitioning scheme is to maximize the number of satisfied slices for all traffic types that receive the RBs necessary to match their size and shape. We can then define a new objective function that seeks jointly maximizing the number of satisfied slices for all traffic types:

$$\max \sum_{\forall s_i \in S_n} [\alpha_i \cdot H_i(K_i)] + \sum_{\forall s_i \in S_p} \left[ \alpha_i \cdot \frac{1}{\lfloor T_w/T_p^i \rfloor} \sum_{\forall t_z \in T_0} H_i(K_i) \right] + \sum_{\forall s_i \in S_a} \left[ \alpha_i \cdot \frac{1}{T_w} \sum_{l=1}^{T_w} H_i(K_i) \right] \tag{24}$$

This objective function is subject to the shape constraints specified in (17), (19)-(20) and (22)-(23). Eq. (24) introduces a priority factor $\alpha_i$ for each slice $s_i$. This factor is used to prioritize slices supporting more critical applications if it is not possible to satisfy all the slices (i.e. to match their size and shape requirements). In this study, we consider the same priority for all slices supporting the same traffic type. The highest priority is for slices supporting deterministic aperiodic traffic, and the lowest one for those supporting non-deterministic traffic. We can then establish that: $\alpha_i > \alpha_j > \alpha_h$ for slices $s_i \in S_a$, $s_j \in S_p$, and $s_h \in S_n$. If all slices are satisfied, the partitioning scheme distributes any available RBs among the slices. It is also possible that RBs remain unassigned even if not all slices are satisfied. This can occur, for example, if available RBs cannot contribute to match the size and (especially) the shape of any unsatisfied slice. The partitioning scheme assigns any available RBs to existing slices in order to improve the QoS they provide. The RBs are distributed taking into account how an increase of the slice's size can impact the QoS provided by the slice[1]. To this aim, we define the following satisfaction function for each slice $s_i \in S_p$ with $K_i$ RBs:

$$Z_i(K_i) = \begin{cases} 0, & \text{if } K_i < K_i^{\text{size}} \\ \dfrac{2}{1 + e^{c(K_i - K_i^{\text{size}})}} - 1, & \text{if } K_i^{\text{size}} \le K_i < K_i^{\text{max}} \\ 1, & \text{if } K_i \ge K_i^{\text{max}} \end{cases} \tag{25}$$

where

$$c = \frac{\ln(0.001)}{K_i^{\text{max}} - K_i^{\text{size}}} \tag{26}$$

and $K_i^{\text{max}}$ represents the size of the slice at which the maximum possible satisfaction is reached. $K_i^{\text{max}}$ is computed using equation (3) with $SINR_u$ equal to the 25th-percentile of the experienced SINR[2]. Figure 7 shows an example of the function $Z_i(K_i)$. The figure shows that the satisfaction of a slice is null if its size requirement is not guaranteed (i.e., $K_i < K_i^{\text{size}}$). From $K_i^{\text{size}}$, the satisfaction increases exponentially with the number of RBs $K_i$ reserved for the slice until $K_i^{\text{max}}$. Adding more RBs from $K_i^{\text{max}}$ does not improve the satisfaction of the slice.

[1] Adding more RBs to a slice does not negatively impact the slice's shape, and hence the capability of the slice to support the latency requirements of the nodes it serves.

[2] $K_i^{\text{size}}$ is computed using equation (3) with $SINR_u$ equal to the average SINR.

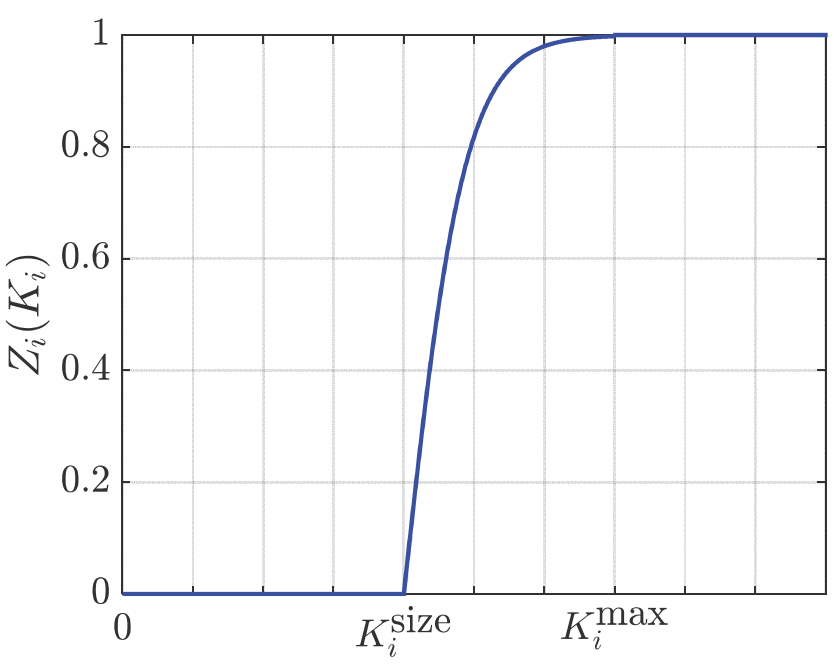

**Figure 7.** Satisfaction function $Z_i(K_i)$.

The partitioning scheme distributes any unassigned RBs with the objective to maximize the sum of the satisfaction perceived by the slices for all traffic types. This is expressed with the following second objective function:

$$\max \sum_{\forall s_i \in S_n} [\alpha_i \cdot Z_i(K_i)] + \sum_{\forall s_i \in S_p} \left[ \alpha_i \cdot \frac{1}{\lceil T_w/T_p^i \rceil} \sum_{\forall t_z \in T_0} Z_i(K_i) \right] + \sum_{\forall s_i \in S_a} \left[ \alpha_i \cdot \frac{1}{T_w} \sum_{l=1}^{T_w} Z_i(K_i) \right] \quad (27)$$

This second objective function is obtained using a similar approach to that used to derive (24). It is also subject to the slices' shape constraints specified in (17), (19)-(20) and (22)-(23). The first and second objective functions ((24) and (27) respectively) can be merged into a single objective function that establishes:

$$\begin{aligned} \max & \sum_{\forall s_i \in S_n} [\alpha_i \cdot H_i(K_i)] + \sum_{\forall s_i \in S_p} \left[ \alpha_i \cdot \frac{1}{\lceil T_w/T_p^i \rceil} \sum_{\forall t_z \in T_0} H_i(K_i) \right] + \sum_{\forall s_i \in S_a} \left[ \alpha_i \cdot \frac{1}{T_w} \sum_{l=1}^{T_w} H_i(K_i) \right] \\ & + \frac{1}{\omega} \cdot \left\{ \sum_{\forall s_i \in S_n} [\alpha_i \cdot Z_i(K_i)] + \sum_{\forall s_i \in S_p} \left[ \alpha_i \cdot \frac{1}{\lceil T_w/T_p^i \rceil} \sum_{\forall t_z \in T_0} Z_i(K_i) \right] + \sum_{\forall s_i \in S_a} \left[ \alpha_i \cdot \frac{1}{T_w} \sum_{l=1}^{T_w} Z_i(K_i) \right] \right\} \end{aligned} \quad (28)$$

The terms in (28) corresponding to the distribution of unassigned RBs (initially (27)) are weighted by $1/\omega$, where $\omega$ is such that $0 < 1/\omega << 1$. This is to prioritize maximizing the number of slices that satisfy their size and shape demand when assigning RBs. Eq. (28) can be expressed as:

$$\max \sum_{\forall s_i \in S_n} [\alpha_i \cdot U_i(K_i)] + \sum_{\forall s_i \in S_p} \left[ \alpha_i \cdot \frac{1}{\lceil T_w/T_p^i \rceil} \sum_{\forall t_z \in T_0} U_i(K_i) \right] + \sum_{\forall s_i \in S_a} \left[ \alpha_i \cdot \frac{1}{T_w} \sum_{l=1}^{T_w} U_i(K_i) \right] \quad (29)$$

where $U_i(K_i)$ is a utility function defined as:

$$U_i(K_i) = H_i(K_i) + \frac{1}{\omega} \cdot Z_i(K_i) \quad (30)$$

The proposed RAN partitioning scheme is then designed to solve the following optimization problem:

$$\text{o.f.:} \max \sum_{\forall s_i \in S_n} [\alpha_i \cdot U_i(K_i)] + \sum_{\forall s_i \in S_p} \left[ \alpha_i \cdot \frac{1}{\lceil T_w/T_p^i \rceil} \sum_{\forall t_z \in T_0} U_i(K_i) \right] + \sum_{\forall s_i \in S_a} \left[ \alpha_i \cdot \frac{1}{T_w} \sum_{l=1}^{T_w} U_i(K_i) \right] \quad (31)$$

$$\text{s.t.:} \sum_{t=1}^{T_w} L_{i,t} = K_i, \quad \forall s_i \in S_n \quad (32)$$

$$\sum_{t=t_z}^{t_z+D_i-1} L_{i,t} = K_i, \forall t_z \in T_0 \;\&\; t_z \leq T_w\text{-}(D_i\text{-}1), \forall s_i \in S_p \quad (33)$$

$$\sum_{t=t_z}^{T_w} L_{i,t} + \sum_{t=1}^{\text{mod}\left(\frac{t_z+D_i-1}{T_w}\right)} L_{i,t} = K_i, \quad \forall t_z \in T_0 \;\&\; t_z > T_w\text{-}(D_i\text{-}1), \forall s_i \in S_p \quad (34)$$

$$\sum_{t=l}^{l+D_i-1} L_{i,t} = K_i, \quad \forall l \in [1, T_w\text{-}(D_i\text{-}1)], \forall s_i \in S_a \quad (35)$$

$$\sum_{t=l}^{T_w} L_{i,t} + \sum_{t=1}^{\text{mod}\left(\frac{l+D_i-1}{T_w}\right)} L_{i,t} = K_i, \quad \forall l \in [T_w\text{-}(D_i\text{-}1)+1, T_w], \forall s_i \in S_a \quad (36)$$

$$\sum_{\forall s_i \in S} L_{i,t} \leq N_{RB}, \forall t \in [1, T_w] \quad (37)$$

$$K_i \leq K_i^{\max}, \forall s_i \in S \quad (38)$$

$$L_{i,t} \in \{0, \mathbb{Z}^+\}, \forall t \in [1, T_w], \forall s_i \in S \quad (39)$$

Eq. (32)-(36) establish the slices' shape requirements for non-deterministic, deterministic periodic, and deterministic aperiodic traffic. Constraint (37) establishes that the number of RBs reserved for all the slices per slot is bounded by the amount of available RBs per slot. The constraint (38) establishes that any slice $s_i$ will not receive more than $K_i^{\max}$ RBs since its satisfaction will not further increase. The constraint in (39) establishes that all possible $L_{i,t}$ solutions must be non-negative integers. It should be noted that different criteria can be applied to select the values of the weight factors or the $Z_i(K_i)$ function for example.

The partitioning problem defined in (31)-(39) is non-linear because the utility function $U_i(K_i)$ defined in (30) is a non-linear function. Powerful methods are available to solve this kind of problems (such as genetic algorithms). However, the computational complexity is usually greater than that of linear optimization problems. In this context, we propose to approximate it with a linear function $U_i'(K_i)$ that is defined as follows:

$$U_i'(K_i) = m_i\, K_i + C_i, \qquad K_i^{\text{size}} \leq K_i \leq K_i^{\text{max}} \tag{40}$$

where

$$m_i = \frac{U_i(K_i^{\text{max}}) - U_i(K_i^{\text{size}})}{K_i^{max} - K_i^{\text{size}}} \tag{41}$$

and $C_i$ is a constant that depends on the variables $K_i^{\text{size}}$ and $K_i^{\text{max}}$.

Figure 8 represents $U_i(K_i)$ and $U_i'(K_i)$ for different values of $1/\omega$. The figure shows that $U_i(K_i)$ can be approximated by the linear function $U_i'(K_i)$ for small values of $1/\omega$. This approximation is valid since we use a value of $1/\omega$ equal to 0.001 (see equation (30)). This value is chosen to prioritize maximizing the number of slices that satisfy their size and shape demand when assigning RBs.

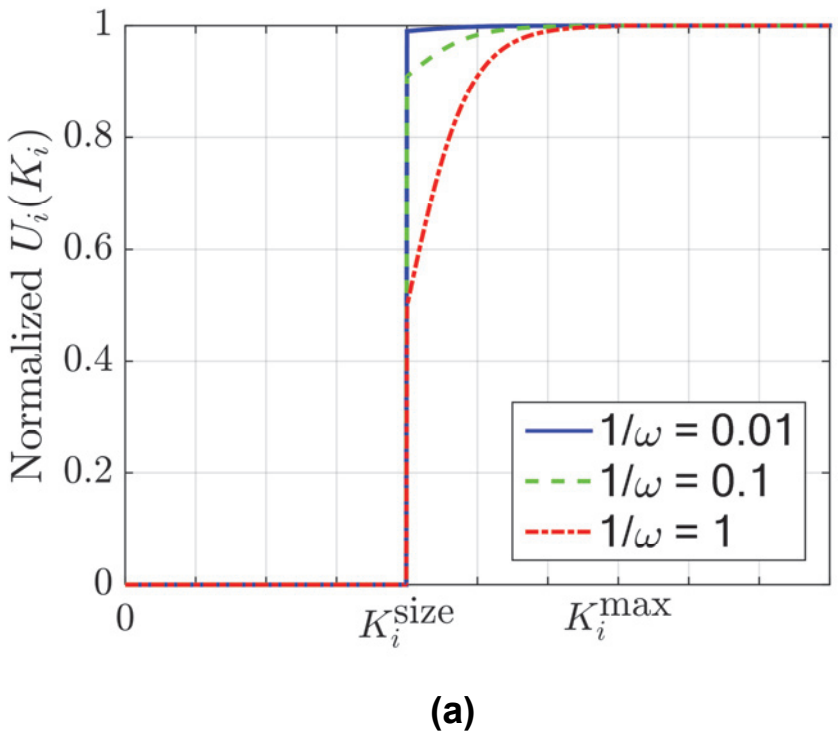

(a)

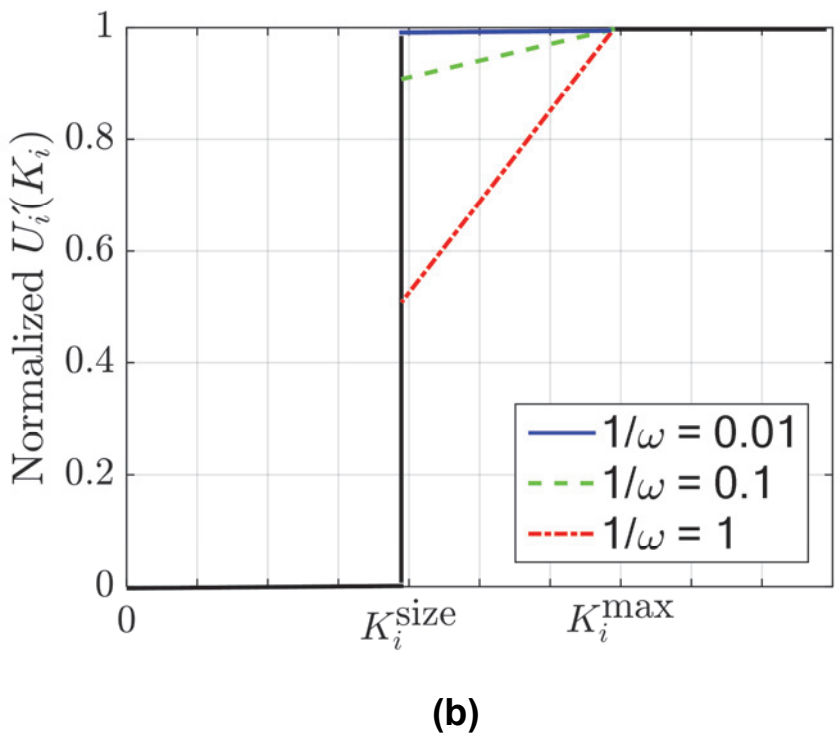

(b)

**Figure 8. Representation of (a) the normalized $U_i(K_i)$ function and (b) its linear approximation $U_i'(K_i)$, for different values of $1/\omega$.**

Using (40), the objective function in (31) can be approximated by:

$$\max \sum_{\forall s_i \in S_n} [\alpha_i \cdot m_i\, K_i] + \sum_{\forall s_i \in S_p} \left[\alpha_i \cdot \frac{1}{\lfloor T_w/T_p^i \rfloor} \sum_{\forall t_z \in T_0} m_i\, K_i\right] + \sum_{\forall s_i \in S_a} \left[\alpha_i \cdot \frac{1}{T_w} \sum_{l=1}^{T_w} m_i\, K_i\right] \tag{42}$$

subject to:

$$K_i^{\text{size}} \leq K_i \leq K_i^{\text{max}}, \qquad \forall s_i \in S \tag{43}$$

It should be noted that $C_i$ is removed from (42) because it is an independent variable that does not impact the result of the maximization of the function in (42). It should also be noted that (42) is a separable function. A separable function is a function where each term is a function of a single variable. In this case, the function is separable into a sum of functions of individual variables [35]. We use this property and the slices' shape requirements in (32)-(36) to express each term in (42) as:

$$\sum_{\forall s_i \in S_n} [\alpha_i \cdot m_i\, K_i] = \sum_{\forall s_i \in S_n} \left[\alpha_i \cdot m_i \cdot \sum_{t=1}^{T_w} L_{i,t}\right] \tag{44}$$

$$\sum_{\forall s_i \in S_p} \left[\frac{\alpha_i}{\lfloor T_w/T_p^i \rfloor} \sum_{\forall t_z \in T_0} m_i\, K_i\right] = \sum_{\forall s_i \in S_p} \left[\frac{\alpha_i \cdot m_i}{\lfloor T_w/T_p^i \rfloor} \sum_{\forall t_z \in T_0} \left(\sum_{t=t_z}^{t_z+D_i-1} L_{i,t}\right)\right] = \sum_{\forall s_i \in S_p} \left[\frac{\alpha_i \cdot m_i}{\lfloor T_w/T_p^i \rfloor} \sum_{t=1}^{T_w} L_{i,t}\right] \tag{45}$$

$$\sum_{\forall s_i \in S_a} \left[\alpha_i \cdot \frac{1}{T_w} \sum_{l=1}^{T_w} m_i\, K_i\right] = \sum_{\forall s_i \in S_a} \left[\alpha_i \cdot m_i \cdot \frac{1}{T_w} \sum_{l=1}^{T_w} \left(\sum_{t=l}^{l+D_i-1} L_{i,t}\right)\right] = \sum_{\forall s_i \in S_a} \left[\alpha_i \cdot m_i \cdot \frac{1}{T_w} \sum_{t=1}^{T_w} D_i L_{i,t}\right] \tag{46}$$

We can then obtain an integer linear optimization problem that is equivalent to the problem defined in (31)-(39). This linear problem is then defined as:

$$\text{o.f.:}\max \sum_{\forall s_i \in S_n} \left[\alpha_i \cdot m_i \cdot \sum_{t=1}^{T_w} L_{i,t}\right] + \sum_{\forall s_i \in S_p} \left[\frac{\alpha_i \cdot m_i}{\lfloor T_w/T_p^i \rfloor} \sum_{t=1}^{T_w} L_{i,t}\right] + \sum_{\forall s_i \in S_a} \left[\alpha_i \cdot m_i \cdot \frac{1}{T_w} \sum_{t=1}^{T_w} D_i L_{i,t}\right] \tag{47}$$

$$\text{s.t.: } K_i^{\text{size}} \leq \sum_{t=1}^{T_w} L_{i,t} \leq K_i^{\max}, \forall s_i \in S_n \quad (48)$$

$$K_i^{\text{size}} \leq \sum_{t=t_z}^{t_z+D_i-1} L_{i,t} \leq K_i^{\max}, \quad (49)$$
$$\forall t_z \in T_0, \& \; t_z \leq T_w\text{-}(D_i\text{-}1), \forall s_i \in S_p$$

$$K_i^{\text{size}} \leq \sum_{t=t_z}^{T_w} L_{i,t} + \sum_{t=1}^{\text{mod}\left(\frac{t_z+D_i-1}{T_w}\right)} L_{i,t} \leq K_i^{\max}, \quad (50)$$
$$\forall t_z \in T_0 \; \& \; t_z > T_w\text{-}(D_i\text{-}1), \forall s_i \in S_p$$

$$K_i^{\text{size}} \leq \sum_{t=l}^{l+D_i-1} L_{i,t} \leq K_i^{\max}, \quad (51)$$
$$\forall l \in [1, T_w\text{-}(D_i\text{-}1)], \forall s_i \in S_a$$

$$K_i^{\text{size}} \leq \sum_{t=l}^{T_w} L_{i,t} + \sum_{t=1}^{\text{mod}\left(\frac{l+D_i-1}{T_w}\right)} L_{i,t} \leq K_i^{\max}, \quad (52)$$
$$\forall l \in [T_w\text{-}(D_i\text{-}1) + 1, T_w], \forall s_i \in S_a$$

$$\sum_{\forall s_i \in S} L_{i,t} \leq N_{RB}, \forall t \in [1, T_w] \quad (53)$$

$$L_{i,t} \in \{0, \mathbb{Z}^+\}, \forall t \in [1, T_w], \forall s_i \in S \quad (54)$$

The partitioning problem defined in (47)-(54) is now an integer linear optimization problem.

## VII. REFERENCE SCHEME

This paper presents a RAN slicing solution that includes novel schemes for the creation of RAN slices and the partitioning of radio resources. An important novelty of the proposed solution is that it utilizes a novel latency-based slice descriptor to improve the capacity of RAN slicing to support latency-sensitive or time-critical services. It is important noting that, to the authors' knowledge, this study is the first to propose latency-based slice descriptors and embed latency in the design of RAN slicing. To date, existing RAN slicing solutions are generally designed as a function of the number of resources assigned to each slice. The review in Section III showed that several relevant contributions (e.g. [20], [25]) define RAN slicing schemes that seek to maximize the sum of the utility obtained by all the slices. These proposals define utility functions that depend on the number of RBs assigned to each slice. The performance of the RAN slicing solution proposed in this study is therefore compared against a utility-based reference scheme. The utility function is defined as a function of the number of RBs assigned to each slice, and the reference scheme seeks to maximize the sum of the utility obtained by all the slices. The reference scheme implemented in this study solves then the following optimization function:

$$\max \sum_{\forall s_i \in S} [\alpha_i \cdot U_i(K_i)] \quad (55)$$

Different utility functions have been proposed in the literature. The objective of this study is not to investigate which is the best utility function. Instead, this study focuses on demonstrating the advantages and gains achieved with RAN slicing solutions that embed latency in their design (and show how to embed it). For a fair comparison, the reference scheme uses the same utility function as (30), and solves then the following optimization problem to determine the number of RBs $K_i$ that are reserved for each slice $s_i$:

$$\text{o.f.: } \max \sum_{\forall s_i \in S} [\alpha_i \cdot U_i(K_i)] \quad (56)$$

$$\text{s.t.: } \sum_{\forall s_i \in S_n} K_i + \sum_{\forall s_i \in S_p} \frac{T_w}{T_p^i} K_i + \sum_{\forall s_i \in S_a} \frac{T_w}{D_i} K_i \leq N_{RB} T_w \quad (57)$$

$$K_i \leq K_i^{\max}, \quad \forall s_i \in S \quad (58)$$

$$K_i \in \{0, \mathbb{Z}^+\}, \quad \forall s_i \in S \quad (59)$$

where $\alpha_i$ is the priority factor of the slice $s_i$. The optimization problem for the reference scheme is similar to that defined in (31), but focuses on optimizing the number of RBs assigned to each slice. Constraint in (57) establishes that the total number of RBs reserved for all the slices is limited to the total number of RBs available in an allocation window (i.e. $N_{RB} T_w$). Constraint (58) establishes that any slice $s_i$ cannot receive more than $K_i^{\max}$ RBs. Following (59), $K_i$ is a non-negative integer.

Equation (57) represents the sum of RBs reserved for all slices of all traffic types. Such part is represented by a sum of three terms. The first term in (57) is the total number of RBs reserved for slices supporting non-deterministic traffic. It is equal to:

$$\sum_{\forall s_i \in S_n} K_i \quad (60)$$

The second term in (57) represents the total number of RBs reserved for slices supporting deterministic periodic traffic. In this case, the size of a slice is defined as the number of RBs that must be reserved within the transmission period to satisfy the rate required by the application. The number of RBs reserved for a slice $s_i \in S_p$ within an allocation window is then given by $(T_w/T_p^i) \cdot K_i$, where $T_w/T_p^i$ represents the ratio between the duration of the allocation window ($T_w$) and the duration of the transmission period for slice $s_i$ ($T_p^i$). The total number of RBs reserved for slices supporting deterministic periodic traffic is then computed as:

$$\sum_{\forall s_i \in S_p} \frac{T_w}{T_p^i} K_i \quad (61)$$

The third term in (57) represents the total number of RBs reserved for slices supporting deterministic aperiodic traffic. In this case, the size of a slice is defined as the number of RBs that must be reserved from the time a packet is generated until the transmission deadline. This time period has a duration equal to $D_i$. The number of RBs reserved for a slice $s_i \in S_a$ is equal to $(T_w/D_i) \cdot K_i$, where $T_w/D_i$ represents the ratio between the duration of the allocation window ($T_w$) and the transmission deadline ($D_i$). The total number of RBs reserved for slices supporting deterministic aperiodic traffic is then computed as:

$$\sum_{\forall s_i \in S_a} \frac{T_w}{D_i} K_i \quad (62)$$

We can linearize the optimization problem in (56). To this aim, the number $K_i$ of RBs reserved for slice $s_i$ can be expressed as:

$$K_i = \sum_{r=0}^{K_i^{\max}} r \cdot x_{i,r} \quad (63)$$

where $x_{i,r}$ is a binary variable equal to one if $r$ RBs are allocated to the slice $s_i$, and equal to 0 otherwise. Similarly, the utility value achieved by slice $s_i$ with $K_i$ RBs can be expressed as:

$$U_i(K_i) = \sum_{r=0}^{K_i^{\max}} U_i(r) \cdot x_{i,r} \quad (64)$$

Using (63) and (64), the optimization problem of the reference scheme can be expressed as a linear optimization problem:

$$\text{o.f.:} \max \sum_{\forall s_i \in S} \left[ \alpha_i \cdot \sum_{r=0}^{K_i^{\max}} U_i(r) \cdot x_{i,r} \right] \quad (65)$$

$$\text{s.t.:} \sum_{\forall s_i \in S_n} \sum_{r=0}^{K_i^{\max}} r \cdot x_{i,r} + \sum_{\forall s_i \in S_p} \frac{T_w}{T_p^i} \sum_{r=0}^{K_i^{\max}} r \cdot x_{i,r} + \sum_{\forall s_i \in S_a} \frac{T_w}{D_i} \sum_{r=0}^{K_i^{\max}} r \cdot x_{i,r} \le N_{RB} T_w \quad (66)$$

$$\sum_{r=0}^{K_i^{\max}} x_{i,r} \le 1, \qquad \forall s_i \in S \quad (67)$$

$$x_{i,r} \in \{0,1\}, \qquad \forall s_i \in S, \forall r \in [0, K_i^{\max}] \quad (68)$$

Constraint (67) indicates that only one $x_{i,r}$ variable can be equal to 1 for each $s_i \in S$. Constraint (68) establishes that $x_{i,r}$ are binary variables.

## VIII. EVALUATION SCENARIO

The RAN slicing proposal is evaluated using Monte-Carlo simulations in Matlab. The simulation platform models the 5G NR radio interface. This study considers a 5G NR numerology $\mu$ equal to 0. An RB is then 180 kHz wide in frequency and lasts for 1ms. Transmissions utilize one of the Modulation and Coding Schemes shown in Table 3. The MCS is dynamically selected based on the SINR. In particular, transmissions select the MCS with larger TBS size that guarantees a target BLER for the experienced SINR. In this study, we consider a target BLER given by 1-$P_{rel}$ = $10^{-5}$, where $P_{rel}$ is the reliability demanded by the application. The simulation platform models the path loss, shadow fading and small scale fading effects. In particular, it implements the path loss model proposed in [36] for the UMi (Urban Micro) scenario and shown in (69):

$$PL = 22.0log10(d) + 28 + 20log10(f_c) \text{ with } f_c\text{=2 GHz} \quad (69)$$

The shadow fading is modelled using a log-normal random distribution with mean equal to 0 dB and standard deviation equal to 3 dB [36]. Rayleigh distribution with zero mean and variance equal to one is considered for the small scale fading. The main communication simulation parameters are summarized in Table 4.

**Table 4. Main simulation parameters**

| Parameter | Value |
|---|---|
| Cell radius | 120 m |
| 5G NR numerology ($\mu$) | 0 |
| Min. distance among BS and nodes | 10 m |
| Transmit power per RB | 16.6 dBm |
| Antenna gain at the BS | 5 dBi |
| Power spectral density of noise | -174 dBm/Hz |
| Noise figure | 9 dB |
| Fading model | Rayleigh $\sim\mathcal{CN}(0,1)$ |
| Shadowing model | Lognormal |
| Shadowing standard deviation | 3 dB |
| Available resources per slot ($N_{RB}$) | 100 RBs |
| Allocation window length ($T_w$) | 10 slots |

We simulate a scenario emulating an industrial plant that is covered by a single 5G NR cell with 120m radius. Nodes in the plant implement different industrial applications with varying communication requirements. We consider scenarios S1, S2 and S3 with 5, 7 and 9 different industrial applications respectively that are selected from Table 5. This table shows the communication requirements for each application following [3]. For each simulation, we randomly select the executed applications from Table 5. On average, the same number of applications demanding deterministic periodic traffic, deterministic aperiodic traffic and non-deterministic traffic is simulated for each scenario. The number of nodes simulated per application follows a Poisson distribution with average $M$ from Table 5. Nodes are homogeneously distributed within the industrial plant. Increasing the number of applications coexisting in the scenario augments the network load and the demand for RBs. The number of RBs is maintained constant and equal to $N_{RB}$ RBs per slot in all the scenarios. As shown in Table 5, this study considers the highest priority for the applications with deterministic aperiodic traffic followed by applications with deterministic periodic traffic. The traffic priorities are utilized only when the number of RBs is not sufficient to satisfy all slices.

Previous studies (such as [15], [20] and [37]) select the allocation window $T_w$ so that the partitioning solution for the current allocation window can be repeated in consecutive allocation windows while satisfying the requirements of the applications supported by the slices. The selection of $T_w$ must consider the requirements and characteristic of the different traffic types. Common industrial applications with deterministic periodic traffic must satisfy that $T_p^i \geq D_i$ [3]. In this case, $T_p^i$ influences what should be the minimum value for the duration of the allocation window. In particular, we must guarantee that at least one periodic transmission occurs within the allocation window (i.e. $T_w \geq T_p^i$). For slices $s_i$ supporting deterministic aperiodic traffic ($s_i \in S_a$), we must guarantee that $T_w \geq D_i$. Slices $s_i$ supporting non-deterministic traffic ($s_i \in S_n$) do not influence the minimum duration of the allocation window. This traffic class demands an average rate $R_i$. This $R_i$ can be satisfied independently of the value of $T_w$. $T_w$ influences the effective transmission rate $R_u^{\text{eff}}$, but it does not impact $R_i$ since the the number of RBs $J_u$ required by node $u$ to achieve $R_i$ are adapted as a function of $T_w$. The minimum duration of the allocation window can then be established considering the conditions identified for deterministic periodic and aperiodic traffic:

$$T_w \geq \max\left\{\max\{T_p^i | \forall s_i \in S_p\}, \max\{D_i | \forall s_i \in S_a\}\right\} \quad (70)$$

**Table 5. QoS requirements of the selected Industry 4.0 applications**

| Traffic class | Application | Latency ($D_i$) | Reliability ($P_{rel}$) | Payload ($b_i$) | Rate ($R_i$) | Average # Nodes ($M$) | Transmission period ($T_p^i$) | Average packet inter-arrival time ($1/\lambda$) | Priority factor ($\alpha_i$) |
|---|---|---|---|---|---|---|---|---|---|
| Deterministic periodic traffic | Printing machine | 2 ms | 0.99999 | 20 bytes | - | 100 | 5 ms | - | 1/10 |
| | Machines coordination | 6 ms | 0.99999 | 1 Kbytes | - | 10 | 10 ms | - | 1/10 |
| | Precise cooperative | 1 ms | 0.99999 | 80 bytes | - | 40 | 2 ms | - | 1/10 |
| Deterministic aperiodic traffic | Emergency stops | 1 ms | - | 40 bytes | - | 30 | - | 60 s | 1 |
| | Closed-loop control events | 2 ms | - | 20 bytes | - | 25 | - | 30 s | 1 |
| | Level alarms for plant asset | 5 ms | - | 32 bytes | - | 15 | - | 10 s | 1 |
| Non-deterministic traffic | Software/firmware updates | - | - | - | 1 Mbps | 12 | - | - | 1/100 |
| | User interaction | - | - | - | 5 Mbps | 4 | - | - | 1/100 |
| | Assets software updates | - | - | - | 1 Mbps | 16 | - | - | 1/100 |

In this study, and without loss of generality, we have set the duration of the allocation window $T_w$ equal to 10 slots[3]. This value has been chosen considering the condition expressed in (70) and the applications selected for this study. These applications are defined in Table 5. The selected applications with deterministic periodic traffic are characterized by transmission periods of 2, 5 and 10 ms. The selected applications with deterministic aperiodic traffic require latency values equal to 1, 2 and 5ms. We can then verify that an allocation window $T_w$ of 10 slots satisfies the condition in (70) for the applications selected in this study and specified in Table 5.

A large number of simulations have been conducted to ensure the statistical accuracy of all the presented results. In particular, we have conducted 1000 simulations (with different seed values) for each scenario configuration (S1, S2 and S3). Each simulation emulates 100.000 allocation windows. The large number of simulations guarantees the statistical accuracy of our results. Figure 9 depicts a flow diagram that illustrates the simulation process. At the start of the simulation, the industrial environment is created: nodes are distributed in the scenario and the industrial applications demanded by the different nodes are selected. Once the environment is created, it is possible to estimate the SINR experienced by each node. A slice is created for each simulated application. All nodes implementing the same application are served by the same slice. Slices are designed once the scenario is created. In this phase, the size and shape of each slice is calculated. To this end, we use information about the communication requirements of the simulated industrial applications and the number of nodes in the scenario. We also consider information about the SINR experienced by each node. After designing the slices, the partitioning problem is formulated and solved. Once a partitioning solution is obtained, transmissions are simulated considering the particular characteristic of the traffic demanded by each industrial application. RBs are then assigned to each active node to carry out its transmission. To this end, 5G NR scheduling allocates resources to the nodes served by a slice. 5G NR introduces the possibility to use grant-based or grant-free scheduling. Grant-based scheduling is used for non-deterministic traffic [39]. In this case, nodes send a scheduling request (SR) to the BS when they require resources, and the BS allocates dedicated RBs using a grant message. This handshaking introduces some latency [40]. Grant-free scheduling is hence considered for deterministic traffic given its stringent latency requirements. In particular, we implement the semi-persistent scheduling (SPS) grant-free scheduling defined in 5G NR [39] for deterministic periodic traffic. With SPS, nodes are assigned dedicated resources from the corresponding slice for a period of time. This approach is suitable for periodic traffic since the assignments can be planned and resources can be utilized efficiently. Such planning is not possible in the case of aperiodic traffic. 5G NR offers the possibility to assign dedicated resources to nodes or shared resources to a group of nodes. We consider the second option for deterministic aperiodic traffic, and implement the grant-free scheduling scheme in [34] that is compliant with the 5G NR standard. Nodes with deterministic aperiodic traffic share all RBs from the corresponding slice. Following [34], nodes randomly select the RBs among available RBs in the slice between the time a packet is generated and its transmission deadline. For each transmission, we log the performance experienced.

[3] This value is also selected in many related studies (e.g. [15], [38]).

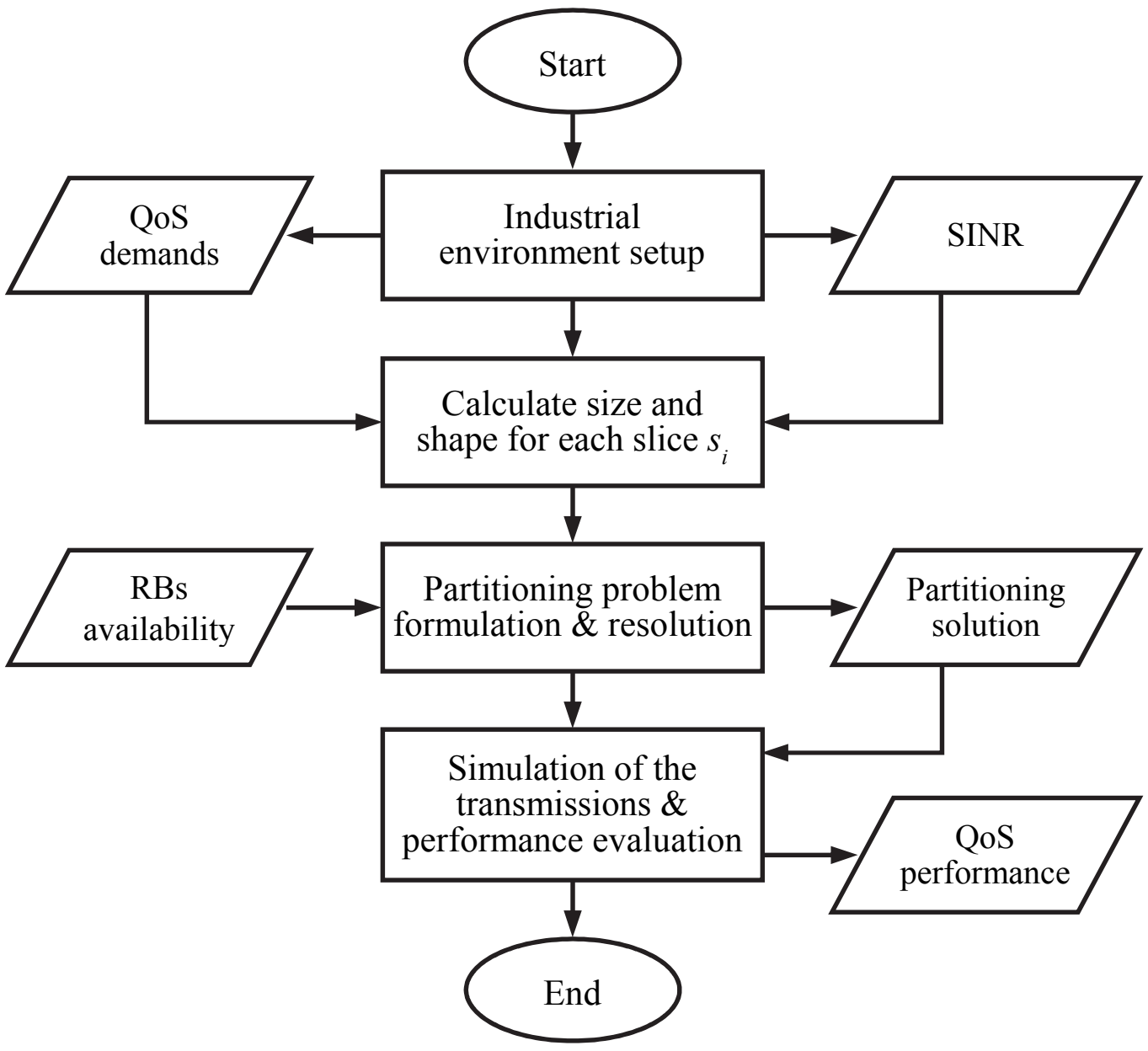

**Figure 9.** **Flow diagram of the simulation process.**

The simulator includes the libraries and functions necessary to solve the optimization problems defined for our RAN slicing proposal and the selected reference scheme. Our RAN slicing proposal defines a non-linear integer programming problem. We utilize a genetic algorithm to solve the problem; in particular, we use the Genetic Algorithm (GA) included in the Matlab Optimization Toolbox. [35] shows that non-linear integer programming problems can be efficiently solved using genetic algorithms. We use the Mixed-Integer Linear Programming (MILP) algorithm provided by the Matlab Optimization Toolbox to solve the integer linear optimization problems defined by the reference scheme (Section VII) and the linear approximation of our proposed partitioning problem (Section VI).

## IX. PERFORMANCE

This section compares the performance achieved with our RAN slicing proposal and with the reference scheme. Our proposal tries first to maximize the number of slices that obtain the RBs necessary to satisfy their size and shape requirements. The proposal then assigns remaining RBs to the slices in order to improve the QoS they can provide. We also analyze the performance obtained when our RAN slicing proposal does not distribute remaining RBs to the slices, i.e. when the partitioning is executed using the objective function in (24) rather than the objective function in (29). The objective of this analysis is to demonstrate that distributing unassigned RBs to slices does not modify the capacity of our RAN slicing proposal to satisfy the data rate and latency requirements (i.e. size and shape) of slices. We refer to this variant as *baseline proposal*.

Figure 10 represents the percentage of slices that receive the RBs necessary to satisfy their size requirements. Figure 11 depicts the percentage of slices that receive the RBs necessary to satisfy their size and shape requirements. The results are depicted for scenarios S1, S2 and S3 with 5, 7 and 9 industrial applications randomly selected per simulation from Table 5. Each application simulates the average number of nodes depicted in Table 5. Increasing the number of applications coexisting in the scenario augments the network load and the demand for RBs. The number of RBs is maintained constant and equal to $N_{RB}$ RBs per slot in all the scenarios. As a result, the probability to satisfy the demand of all nodes decreases with the number of applications in a scenario. Results in Figure 10 and Figure 11 are represented per traffic class and considering all traffic classes together (*Total* in Figure 10 and Figure 11). The first important observation from Figure 10 and Figure 11 is the linear approximation of our partitioning scheme achieves the same results as the original non-linear partitioning solution. This shows that the linear approximation does not reduce the effectiveness of the allocation of RBs to slices achieved with our RAN slicing proposal. Figure 10 and Figure 11 also show that our RAN slicing proposal (whether using the non-linear or linear approximation of the partitioning scheme) always achieves the same percentage of satisfied slices than the baseline proposal variant[4]. This demonstrates that distributing unassigned RBs to slices does not modify the capacity of our proposal to maximize the number of slices that receive the RBs necessary to satisfy their size and shape requirements.

Figure 10 shows that all schemes can satisfy the rate demand of all slices (i.e. their size) when the number of applications and network load is low (i.e. S1). However, this is not possible when the number of applications increases. In this case, more slices must be created, and they all compete for the available bandwidth (and RBs). This bandwidth is maintained constant and equal in this study for all scenarios. Consequently, it is not possible to satisfy all slices with the available bandwidth independently of the RAN slicing solution that is utilized. The percentage of satisfied slices therefore decreases with the number of applications. Figure 10 shows that our RAN slicing proposal achieves equal or higher percentage of slices that achieve their rate (or size) demand than the reference scheme except for deterministic periodic traffic under S2 and S3. This result is independent of whether we utilize the original partitioning proposal or its linear approximation. Our proposal only assigns RBs to slices if they contribute to guarantee both the rate and latency demand (i.e. the slices' size and shape). On the other hand, the reference scheme only takes into account the rate demand (i.e. the size of the slices) to distribute the RBs. This results in that the reference technique assigns more RBs within an allocation window to slices serving deterministic periodic traffic than our RAN slicing proposal (Figure 12). Consequently, the reference technique increases the percentage of slices that satisfy their rate demand for periodic deterministic traffic (Figure 10). This is done at the expense of the slices serving non-deterministic traffic that receive less RBs per allocation window (Figure 12[5]) and achieve lower satisfaction levels (Figure 10).

[4] This does not mean that both schemes assign the same number of RBs per slice and can equally satisfy the nodes served per slice. These results are analyzed in following figures.

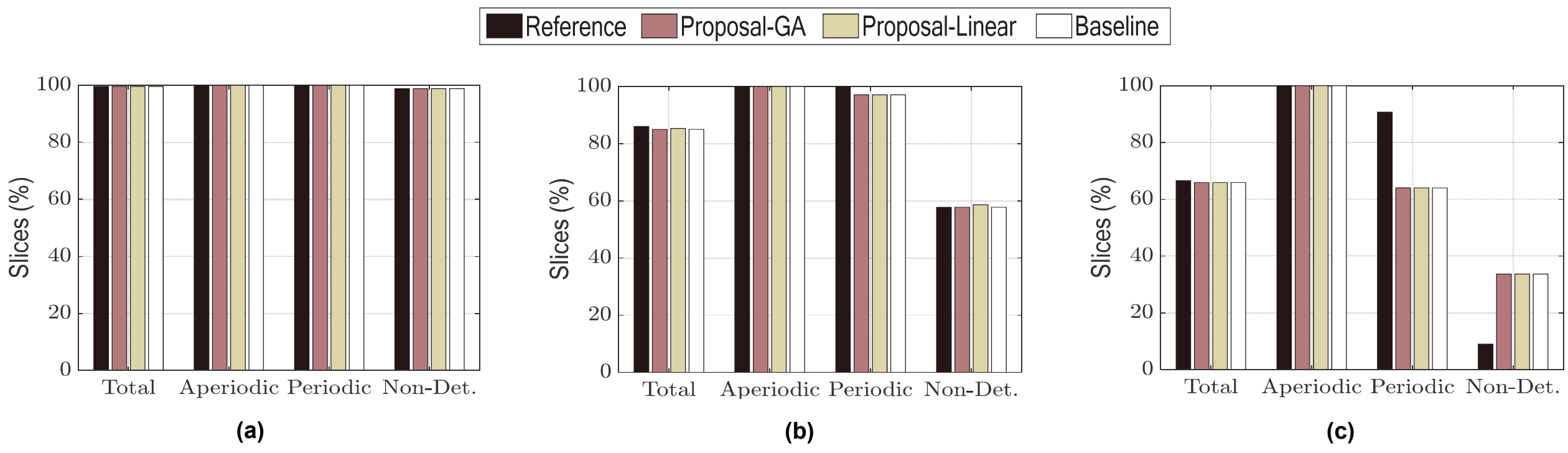

**Figure 10. Percentage of slices that receive the RBs necessary to satisfy their rate demand (i.e. their size). (a) Scenario S1 with 5 industrial applications; (b) Scenario S2 with 7 industrial applications; (c) Scenario S3 with 9 industrial applications.**

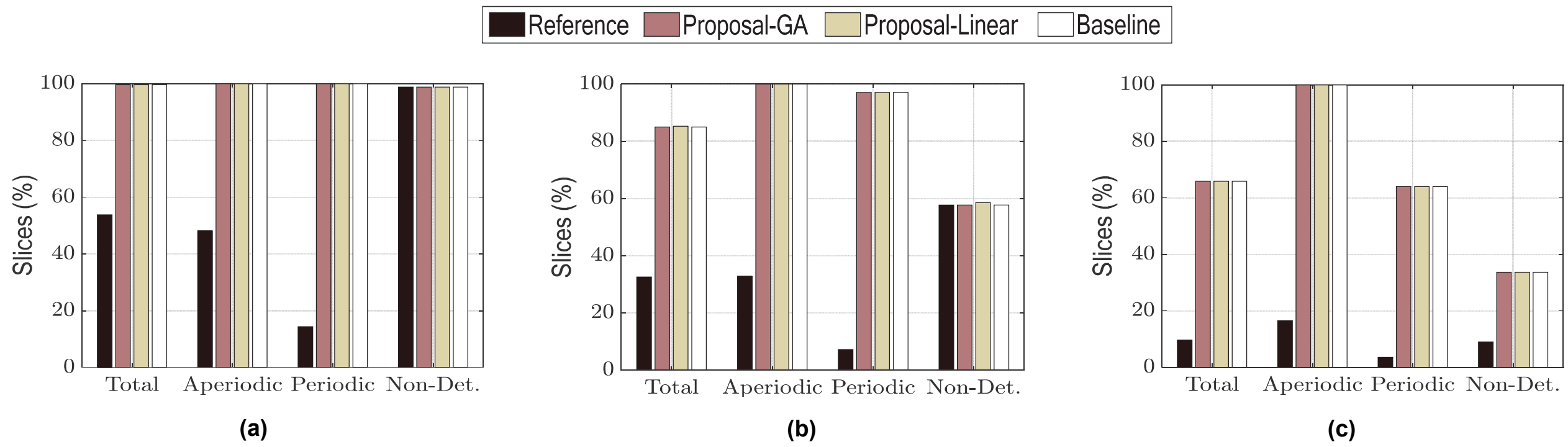

**Figure 11. Percentage of slices that receive the RBs necessary to satisfy their rate and latency demand (i.e. their size and shape). (a) Scenario S1 with 5 industrial applications; (b) Scenario S2 with 7 industrial applications; (c) Scenario S3 with 9 industrial applications.**

Figure 10 showed that the reference scheme can assign the RBs necessary to satisfy the rate demand (i.e. the size) of a large percentage of slices. However, satisfying the rate demand does not imply that the slice has the RBs necessary to satisfy the latency requirements of the nodes it supports. This is actually visible when comparing Figure 10 and Figure 11. For example, Figure 10 shows that the reference scheme can satisfy the rate demand of 85.1% of slices of all traffic types under S2. However, it can only satisfy the rate and latency demand of 30.7% slices in the same scenario. This degradation is particularly relevant when analyzing deterministic periodic traffic. Figure 10 showed that the reference technique can satisfy the rate demand for a larger percentage of slices than our RAN slicing proposal under S2 and S3. However, Figure 11 shows that our RAN slicing proposal significantly outperforms the reference technique when analyzing the percentage of slices that satisfy both their rate and latency demands under S2 and S3. In fact, Figure 11 shows that our RAN slicing solution always outperforms the reference scheme in terms of percentage of slices that receive the RBs necessary to satisfy their rate and latency demand. The figure shows that very similar results are obtained whether using our original partitioning proposal or its linear approximation. As a result, in the rest of this section, we focus on the performance obtained using the original non-linear partitioning scheme. Our proposal is able to satisfy the rate and latency demands of all slices (of all traffic types) under S1. This is not the case for the reference scheme that can only satisfy the rate and latency demands of 48.8% and 3.6% of the slices supporting applications with deterministic aperiodic and periodic traffic respectively[6]. Like in Figure 10, the percentage of slices that receive the RBs necessary to satisfy their rate and latency demand decreases with the number of applications. However, Figure 11 shows that our RAN slicing proposal is able to satisfy a significantly larger percentage of total slices compared to the reference scheme. The differences observed between traffic types are due to the traffic prioritization defined in Section VI. This study considers the highest priority for the applications with deterministic aperiodic traffic followed by applications with deterministic periodic traffic[7]. The traffic prioritization has an impact when the number of RBs is not sufficient to satisfy all slices. For example, Figure 11 shows that our RAN slicing proposal can satisfy the rate and latency demand of all slices with deterministic aperiodic traffic (highest priority) in scenarios with 7 (S2) and 9 (S3) applications. On the other hand, the reference scheme can only satisfy 33.4% and 16.9% of the slices in these two scenarios even if the same traffic priorities are applied. Our proposal achieves similar gains for the slices with deterministic periodic traffic (second priority). In this case, it can satisfy the rate and latency demand of 93.1% and 65.7% of the slices under S2 and S3 with the available RBs. The reference scheme can only satisfy 3.7% and 3.5% of slices for deterministic periodic traffic under S2 and S3. Our proposal also improves the performance achieved for non-deterministic traffic. These results clearly demonstrate that our RAN slicing proposal can better satisfy the applications' rate and latency demands than existing RAN slicing techniques. These techniques generally base the distribution of RBs between slices on the rate demands and ignore the latency. On the other hand, this study has proposed a novel way to create slices and partition the RBs that also takes into account the latency requirements. The results in Figure 11 clearly demonstrate the effectiveness of our proposal.

[5] Again, no significant differences are observed with our RAN slicing proposal using the original partitioning proposal or its linear approximation.

[6] The reference scheme satisfies 100% of the slices for applications with non-deterministic traffic since this traffic does not have latency requirements.

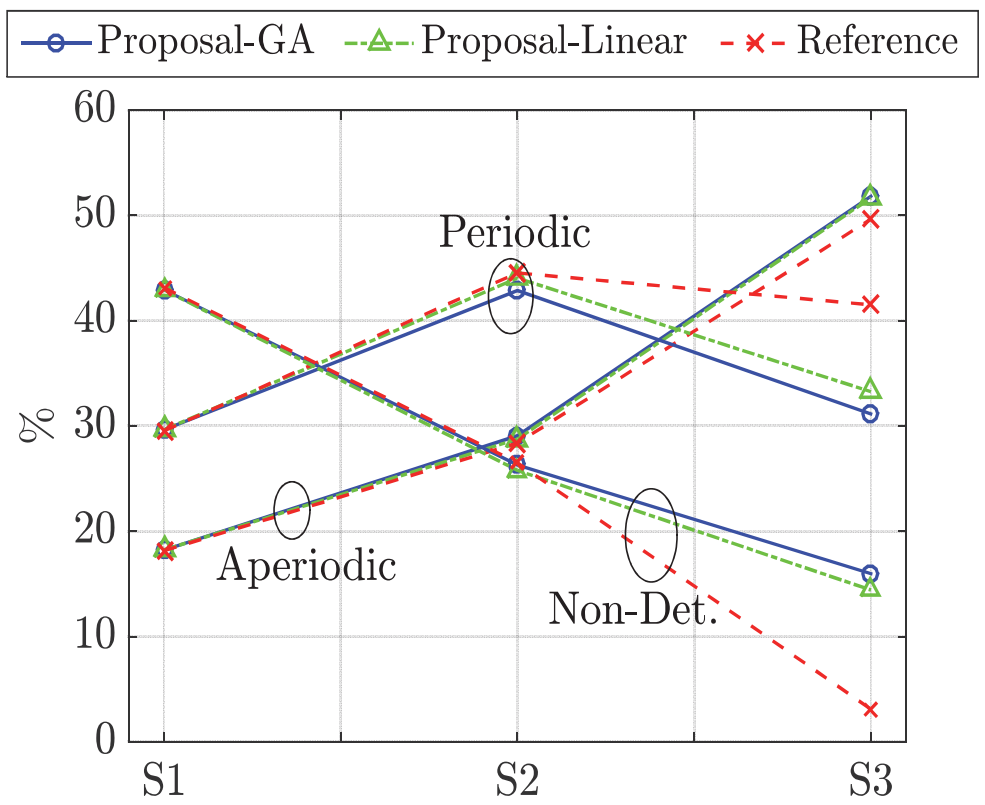

**Figure 12. Percentage of RBs assigned per slice during an allocation window. Results are shown for the three different scenarios.**

Our RAN slicing proposal distributes the RBs with the objective to first maximize the number of slices that satisfy their rate and latency (i.e. size and shape) demands. It then assigns any remaining RBs to improve the QoS slices can provide. Following Section VI, slices cannot receive more than $K_i^{\max}$ RBs, and the size of a slice is denoted $K_i^{\text{size}}$. Figure 13 shows the percentage of slices that receive a number $K_i$ of RBs between $K_i^{\text{size}}$ and $K_i^{\max}$. Figure 13 shows that our proposal allocates more than $K_i^{\text{size}}$ RBs to all slices under S1. In fact, it can even allocate $K_i^{\max}$ RBs to a high percentage of slices (more than 80%) since there are more RBs available than needed by all slices. When the number of applications increases (and hence the network load), the RBs must be distributed among a larger number of slices. Our proposal detects this change, and reduces the percentage of slices that receive $K_i^{\max}$ RBs so that more slices can be satisfied with at least $K_i^{\text{size}}$ RBs. In this case, the percentage of slices that receive only the number of RBs required to satisfy their size $K_i^{\text{size}}$ increases, and the number of slices that receive $K_i^{\max}$ RBs decreases (Figure 13). These results show that our proposal is able to adaptively configure the number of RBs per slice based on the applications' rate and latency demands and on the network load.

Figure 14 compares the average *Successful Transmission Ratio (STR)* achieved with our RAN slicing proposal (using the original non-linear partitioning scheme) and the reference scheme. This metric quantifies the percentage of successful transmissions. A transmission is considered successful if it achieves the QoS level demanded by the corresponding application following Table 5. The QoS requirement of non-deterministic traffic is the data rate demanded by the application. The QoS demand for applications with deterministic (periodic and aperiodic) is defined by the required data rate and the transmission deadline. The results in Figure 14 relate to the capacity of the RAN slicing schemes to create slices with the size and shape necessary to satisfy the rate and latency demands of the applications (Figure 10 and Figure 11). Figure 14 depicts the average *STR* as a function of the number of applications in the scenario for each traffic type. The figure shows that our proposal outperforms the reference scheme in all the scenarios and for all traffic types. Figure 14.a shows that our proposal guarantees the QoS requirements for all deterministic aperiodic transmissions. This is not the case of the reference scheme even if this traffic type has the highest priority in this study[8]. Our proposal guarantees then that any aperiodic transmission can be completed before the deadline established by the application. This is very relevant since this traffic relates to critical events in Industry 4.0 such as emergency stops or failure alarms. Adequately serving this traffic without over-dimensioning the network is important for the future deployment of 5G in factories. The performance achieved with our RAN slicing proposal for deterministic aperiodic traffic is not obtained at the expense of the other traffic types. Figure 14 shows that our proposal outperforms the reference scheme for all traffic types. The performance decreases with the number of applications given the lack of RBs to satisfy all slices and applications under the evaluated conditions (see also Figure 11). However, the degradation is smaller with our RAN slicing proposal than with the reference scheme. Figure 14 shows that our RAN slicing proposal significantly outperforms the reference scheme for deterministic periodic traffic. Figure 12 showed that the reference scheme increases the number of RBs assigned per slice serving deterministic periodic traffic. The reference scheme also augments the percentage of slices with a size that satisfies the applications' rate demand (Figure 10). Despite these results, the reference scheme achieves a significantly lower average STR compared to our RAN slicing proposal (Figure 14). This clearly demonstrates that having more RBs per slice does not directly imply a better capacity to guarantee the latency deadline of deterministic traffic. Our proposal can better guarantee the latency deadlines of deterministic traffic with less RBs per slice than the reference scheme. This is thanks to directly considering the latency demands when creating the slices and partitioning the RBs.

[7] The advantages of our RAN slicing proposal over the reference scheme are maintained with different priorities.

[8] The reference scheme satisfies 84% of the transmissions corresponding to deterministic aperiodic traffic. This percentage is higher than the percentage of slices with the size and shape necessary to satisfy the QoS demands (Figure 9). It should be noted that a slice can satisfy certain nodes with its RBs even if it does not have the shape required for satisfying all the nodes it serves.

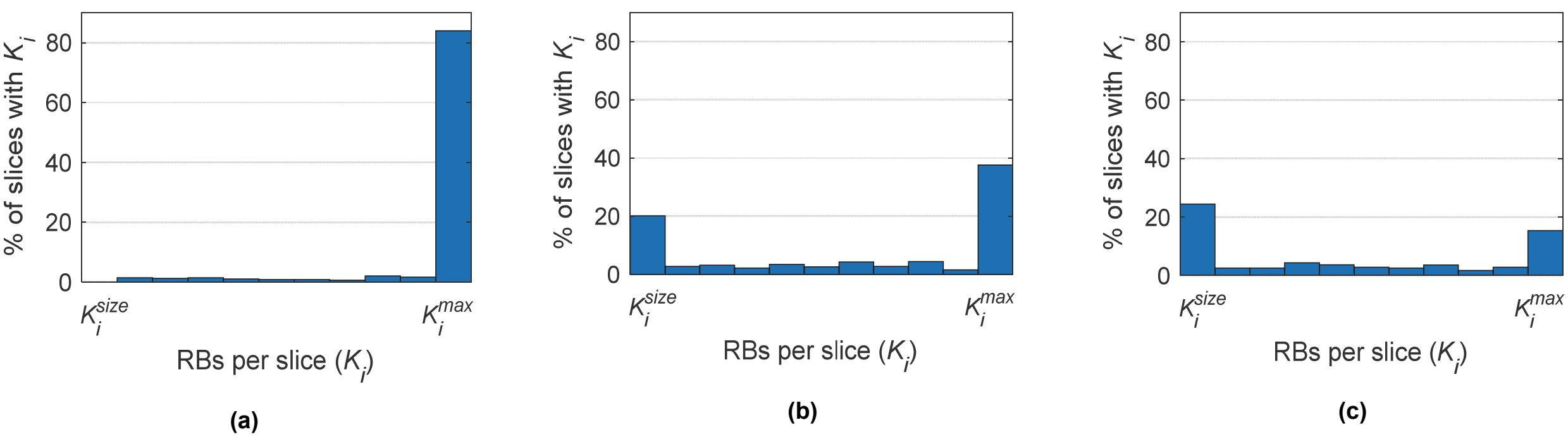

**Figure 13. Number of RBs reserved per slice by our RAN slicing proposal for each scenario: (a) Scenario S1 with 5 industrial applications; (b) Scenario S2 with 7 industrial applications; (c) Scenario S3 with 9 industrial applications.**

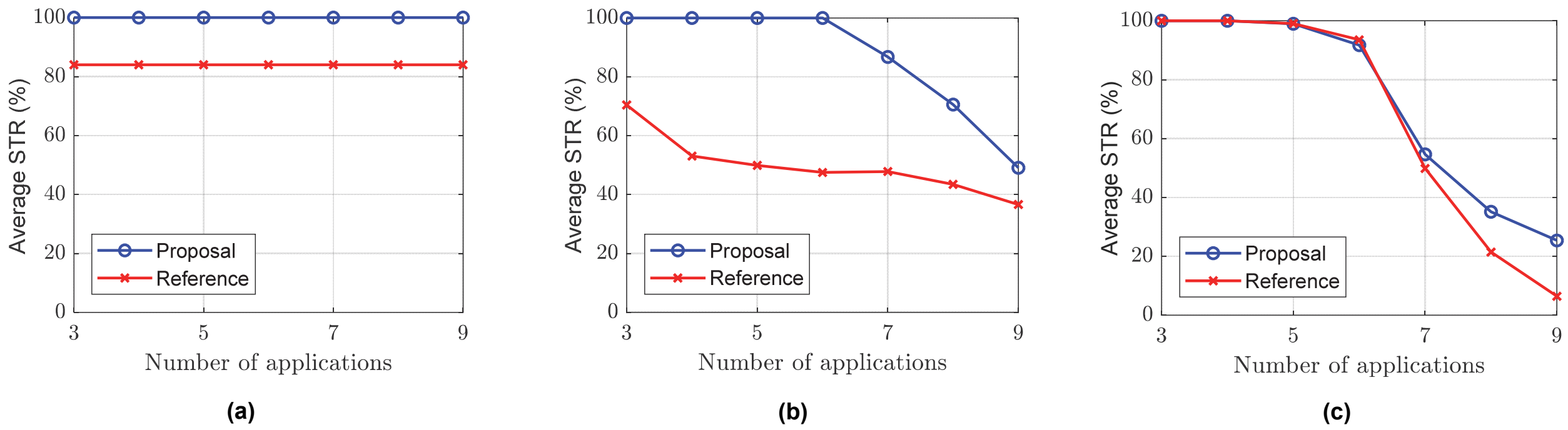

**Figure 14. Average *STR* as a function of the number of applications in the scenario: (a) Deterministic aperiodic traffic; (b) Deterministic periodic traffic; (c) Non-deterministic traffic.**

The performance evaluation is completed with an analysis of the computational cost of our proposed RAN slicing solution. In particular, we focus on the computational cost of the partitioning scheme since this is the module that allocates RBs to slices based on the operating conditions and the traffic types; it is also the module that requires a larger execution time since scheduling schemes are executed in real-time. We compute the computational cost of our original non-linear partitioning scheme and its linear approximation. We utilize a Genetic Algorithm (GA) to solve the non-linear optimization problem and a MILP algorithm to solve its linear approximation; in both cases, we use the Matlab Optimization Toolbox. [35] shows that non-linear integer programming problems can be efficiently solved using genetic algorithms. The computational cost is evaluated using a server with Intel(R) Xeon(R) Gold 6130 processor and a CPU at 2.1GHz. A large number of simulations has been conducted and Table 6 reports the average execution time of our partitioning scheme implemented using the GA and MILP algorithms. The table reports the computational cost for all the considered scenarios. The obtained results show that our partitioning scheme has a low computational cost and can allocate RBs to slices in a short time. In particular, the original non-linear implementation achieves a partitioning solution on average in less than 160 ms while the execution time is reduced to a maximum of 50 ms for the linear implementation. As expected, the execution time increases with the number of applications in the scenario but the execution times are still low. It is also important noting that the partitioning scheme might not be executed that frequently. In particular, it is executed when there are changes in the operating conditions (at the network and traffic level) and it is necessary to re-organize RBs among slices serving different traffic types and number of users. If the conditions do not change, the allocation of RBs to slices is maintained for different allocation windows.

**Table 6. Time-efficiency of the proposed algorithms.**

| Algorithm | S1 | S2 | S3 |
|---|---|---|---|
| Proposal-GA | 151.6 ms | 158.3 ms | 160.4 ms |
| Proposal-Linear | 20.9 ms | 41.3 ms | 49.7 ms |

## X. CONCLUSIONS

This paper has presented a novel latency-sensitive 5G RAN slicing solution. The proposal has been evaluated in Industry 4.0 scenarios with mixed traffic types. This includes applications with deterministic aperiodic, deterministic periodic and non-deterministic traffic. The 5G RAN slicing proposal designs slices and partitions (or allocates) radio resources among slices considering the rate and latency demands of the applications. The study has demonstrated that the proposal improves the capacity of 5G to satisfy the latency requirements of latency-sensitive or time-critical Industry 4.0 applications compared to current solutions based on rate demands. The proposal improves the QoS experienced by all traffic types thanks to a more efficient allocation of radio resources to the slices. The proposed 5G RAN slicing solution has been designed and tested in Industry 4.0 scenarios. However, it could be evolved to support other verticals (e.g. automotive) with latency-sensitive or time-critical applications.